\documentclass[%
 reprint,
 amsmath,amssymb,
 aps,
]{revtex4-2}
\usepackage{graphicx}% Include figure files
\usepackage{dcolumn}% Align table columns on decimal point
\usepackage{bm}% bold math
\usepackage{xcolor}
\usepackage{amsmath}
\usepackage{mathtools}
\usepackage{tabularx}

\makeatletter

\newcommand{\Rmnum}[1]{\expandafter\@slowromancap\romannumeral #1@}
\makeatother
\begin{document}

\title {Tuning exciton complexes in twisted bilayer WSe$_2$ at intermediate misorientation}
\author{Rahul Debnath$^{1\dagger}$, Shaili Sett$^{1\dagger}$, Sudipta Kundu$^{1}$, Rabindra Biswas$^{3}$, Varun Raghunathan$^{3}$, Manish Jain$^{1}$, Arindam Ghosh$^{1,2}$ and Akshay Singh$^{1}$ }
\vspace{1.5cm}
\address{$^1$Department of Physics, Indian Institute of Science, Bangalore 560012, India}
\address{$^2$Centre for Nano Science and Engineering, Indian Institute of Science, Bangalore 560012, India}
\address{$^3$Department of Electrical and Communication Engineering, Indian Institute of Science, Bangalore 560012, India}
\address{$^\dagger$ Contributed equally}

\begin{abstract}
Twist angle modifies the band alignment, screening and interlayer coupling in twisted bilayers of transition metal dichalcogenides. Intermediate misorientation (twist angles $\theta>15^\circ$) bilayers offer a unique opportunity to tune excitonic behaviour within these concurrent physical mechanisms but are seldom studied. In this work, we measure many-body excitonic complexes in monolayer (ML), natural bilayer (BL) and twisted bilayer (tBL) WSe$\mathrm{_2}$. Neutral biexciton (XX) is observed in tBL for the first time, while being undetected in non-encapsulated ML and BL, demonstrating unique effects of disorder screening in twisted bilayers. The XX as well as charged biexciton (XX$\mathrm{^-}$), are robust to thermal dissociation, and are controllable by electrostatic doping. Vanishing of momentum indirect interlayer excitons with increasing electron doping is demonstrated in tBL, resulting from near-alignment of $Q-K$ and $K-K$ valleys. Intermediate misorientation samples offer a high degree of control of excitonic complexes, while offering possibilities of studying exciton-phonon coupling, band-alignment, and screening.
\end{abstract}

\maketitle

\section{\textbf{introduction}}
A moir\'{e} potential results from the relative atomic arrangement, when two van der Waals materials are stacked together at zero or relative twist angles. Twisted bilayers (tBL) of two-dimensional transition metal dichalcogenides (TMDCs) exhibit novel opto-electronic features as a result of the underlying emergent potential. \cite{yu2017moire, tran2020moire}. The rotational alignment between the two layers has emerged as a key degree of freedom, that can control the momentum alignment of the valleys and the hybridization of respective layers \cite{andersenexcitons}. Emergent phenomena such as interlayer (IL) excitons localized by the moir\'{e} potential \cite{tran2019evidence, jin2019observation}, observation of hybridized excitons \cite{ alexeev2019resonantly}, and anti-bunching of moir\'{e} exciton emission  \cite{baek2020highly} have been established. Many-body correlated physics, ranging from moir\'{e} excitons and moir\'{e} trion \cite{brotons2021moir}, Stark effect in excitons \cite{tang2020tuning}, Wigner crystal state \cite{regan2020mott}, and stripe phases \cite{jin2020stripe} have been observed in twisted bilayers of TMDCs.

WSe$_2$ in the 2D limit (with large spin-orbit coupling) is ideal for hosting multi-particle bound states due to reduced dielectric screening of Coulomb interactions, resulting in formation of tightly bound excitons and correlated excitonic states \cite{you2015observation}. WSe$_2$ bandgap changes from a direct K-K transition (in monolayer) to an indirect gap Q-K transition in bilayer (BL) \cite{he2020valley}. The relative band-alignment of Q and K valleys is also sensitive to the rotational alignment between the layers \cite{latini2017interlayer,huang2014probing,liu2014evolution}, making WSe$_2$ an ideal platform for studying the momentum indirect spectral emissions. So far, the studies on excitons in WSe$_2$ and other TMDCs have primarily focused on identification of multi-particle bound states in monolayers (ML) and natural BL \cite{ross2013electrical,li2018revealing,he2014tightly,wang2014exciton}. Established spectral features include bright neutral exciton ($X^{0}$) and trions ($X^{-}$, $X^{+}$), and the dark exciton ($X^{D}$). The dark exciton \cite{lindlau2018role, zhang2015excited, wang2014exciton} has an out-of-plane dipole and usually can-not be spectroscopically observed, unless using a large numerical aperture objective or in-plane magnetic field \cite{chen2018coulomb, liu2019valley, li2019emerging}. A schematic of the formation of bright and dark excitons is shown in Fig. 1(a). Exhaustive investigations involving electrostatic doping, power-dependent photoluminescence (PL) and thermal activation studies have established that emission features at 18 - 25 meV and 47 - 54 meV below $X^{0}$ are the biexciton ($XX$) and charged biexciton ($XX^{-}$) respectively \cite{chen2018coulomb, forste2020exciton, steinhoff2018biexciton, hao2017neutral}. Fig. 1(b) shows a schematic of the formation of these higher exciton complexes. At higher electron doping, doubly negative charged trion ($X^{--}$) has also been observed \cite{barbone2018charge}. Understanding the physical behaviour of multi-particle states with temperature and electrostatic doping is essential to harness the true capability of the twisted artificial systems in opto-electronic applications, and is the main focus of our work.
\begin{figure*}
\includegraphics[width=1\linewidth]{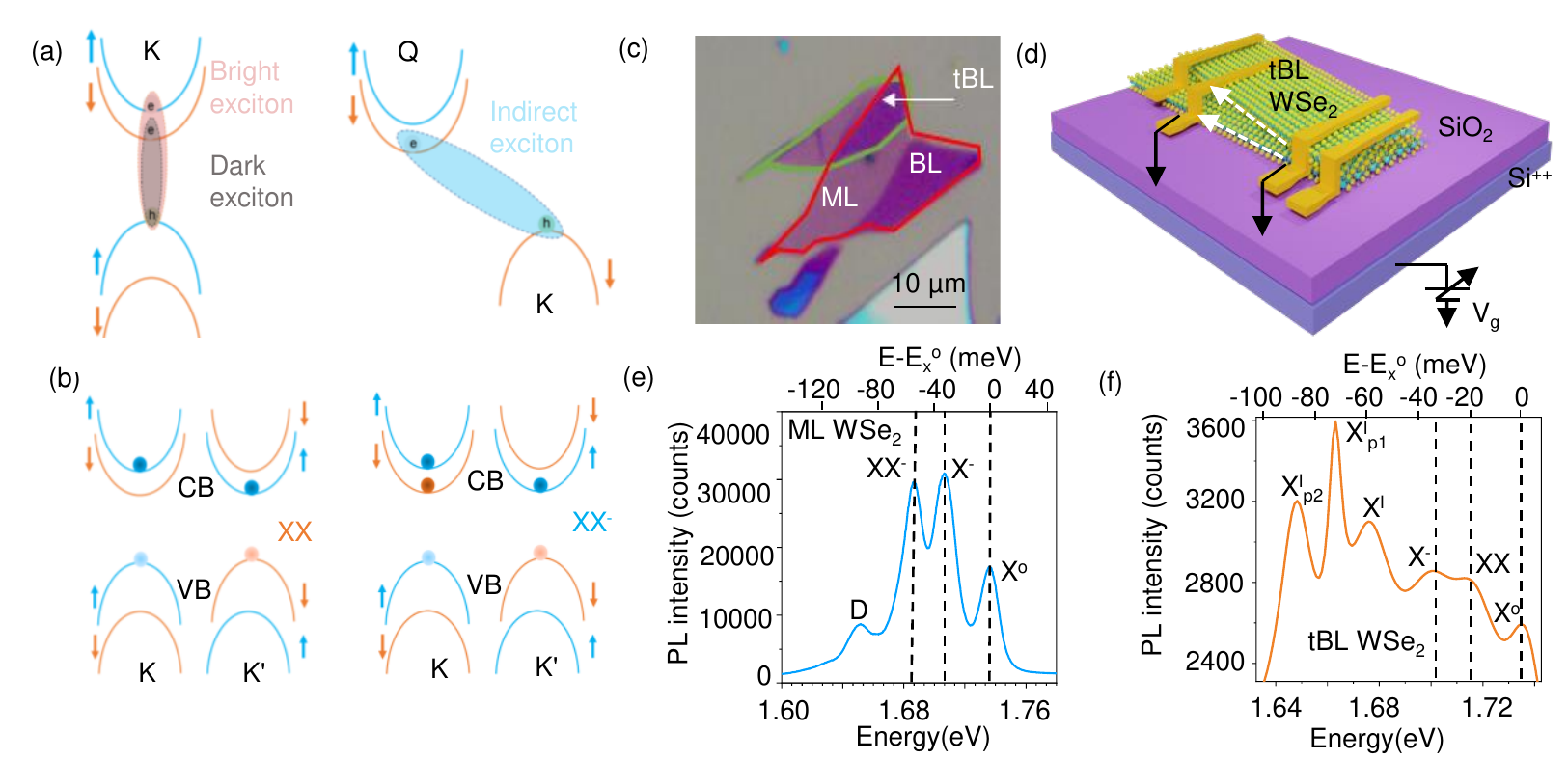}
\caption{\textbf{Bound exciton states, device configuration and PL spectra .} (a) and (b) Schematic of multi-particle bound states in ML and BL . The arrows signify the spin of the charged particle. The electrons are shown as opaque circles while the holes are transparent circles. The schematic shows the bound states of neutral exciton , dark exciton , indirect interlayer exciton, neutral biexciton (XX) and negatively charged biexciton.(c) Optical image of monolayer (ML), bilayer (BL) and twisted bilayer (tBL) regions. (d) Device schematic with electrical configuration. (e) and (f) PL spectra of ML and tBL at temperature 7 K.  }
\end{figure*}

 This work combines PL spectroscopy over a broad temperature range and first principles density functional theory (DFT) analysis to study the interplay of charge, layer coupling and band structure in $\approx$ 20$^{\circ}$ tBL WSe$_2$, as well as its ML and BL counterparts. We observe higher order excitonic complexes at high temperatures in all the WSe$_2$ samples, with majority PL contribution in the high electron doped regime. Our measurements demonstrate the spectral signature of the neutral biexciton in tBL WSe$_2$ up to 180 K, signifying its robustness and stability, and effects of screening. Thermally activated measurements were performed to explore the underlying energetics of the exciton complexes, using which we determine their binding energy and establish their presence in the WSe$_2$ system. Additionally, power dependent PL for $XX$ show super-linear power dependence, further supporting the peak assignment of $XX$. Further, twist angle causes a shift in the energy of phonon-assisted momentum-indirect IL excitons, and these observations are corroborated by DFT calculations. Remarkably, IL exciton in tBL shows asymmetric PL spectra with doping, which is attributed to near-alignment of Q-K and K-K transitions. Overall, our work provides new insights into the charge tunability of the higher-order excitonic complexes in intermediate misorientation tBL WSe$_2$.

\section{\textbf{Results and discussion}}

\textit{\textbf{Device configuration of tBL WSe$_2$:}}  Fig. 1(c) shows the tBL WSe$_2$ heterostructure. The transferred 2D stack has three distinct portions; ML, natural BL (marked as BL) and tBL. The relative rotation between the individual layers is 20$^{\circ}$ $\pm$ 0.5$^{\circ}$ that has been verified optically \cite{debnath2021simple}, and also by second harmonic generation (SHG). Details are given in Supporting Information (SI), section 1. A field-effect transistor (FET) structure is fabricated which is equipped with a global SiO$_2$ back gate for electrostatic doping and the individual 2D layers can be grounded through the metallic contacts (Cr/Au), as shown in the schematic of Fig. 1(d). PL measurements were performed using a 532 nm continuous-wave (CW) laser.

\textit{\textbf{Excitonic complexes in a WSe$_2$ system:}}
The device design enables measurement of PL in three distinct systems under identical temperature and doping conditions. Fig. 1(e) and (f) shows the PL spectrum of ML and tBL WSe$_2$ respectively at 7 K. In ML and BL WSe$_2$ (see SI, section II for PL data on BL), under zero gate bias, we observe three excitonic species: $X^{0}$ at 1.736 eV, $X^{-}$ (34 meV below $X^{0}$), and $XX^{-}$ at 1.685 eV (17 meV below $X^-$) \cite{courtade2017charged,chen2018coulomb, barbone2018charge}. We note that there is significant peak broadening that disallows differentiation between $XX^-$ and $X^{--}$ peaks at high electron doping, and thus we denote the peak as $XX^-$($X^{--}$). Apart from identification of the excitonic peaks from their spectral position, further evidence is provided by thermal activation studies to determine the binding energy (BE), as discussed in the last section of this paper. The lower energy peaks from 1.5 - 1.65 eV in the tBL are attributed to the indirect excitons ($X^{I}$), occurring due to indirect momentum $Q-K$ transition. The energetic position of the $Q$ point in the Brillouin Zone is discussed in SI,section V. Multi-peak Lorentzian fits to the PL spectra for all devices are given in SI, section II. Table 1 provides the peak positions of the various excitons observed in the WSe$_2$ system and compares them with those available in literature. It is to be noted that the absolute peak position for excitonic complexes depends on the sample quality and dielectric environment. However, the relative difference in peak position (binding energy) of the excitonic complexes is unlikely to change between samples. We have observed similar binding energy for excitonic complexes in our samples, compared to those in literature.
\\
\textit{\textbf{Electrostatic tuning of many-body states:}}
We now investigate the gate-voltage dependence of the PL spectrum. Figures 2(a) and 2(b) show doping-dependent color maps of the PL spectra in ML and BL WSe$_2$ respectively, with 100 ${\mu}$W excitation power. In ML, as electrons are removed from the sample by applying a negative gate voltage,  PL intensity of $X^{0}$ increases as the sample becomes more charge neutral, whereas  PL intensity of $XX^-$rapidly decreases (see Fig. 2(a)). The $XX^{-}$ peak exists only between the crossover from neutral to the n-doped regime, implying that it is indeed negatively charged (formed from a bound state of $X^{-}$ and $X^D$).  Electron doping through electrostatic back-gate voltage ($V_g$) ($0 V < V_g < 30 V$) leads to conversion of $X^{0}$ to $X^{-}$. Enhancement of the $XX^{-}$ peak upon further electron doping ($V_g$ $<$ 60 V) is observed, which is possibly due to emergence of the $X^{--}$ \cite{li2019emerging, barbone2018charge}. Fig. 2(c) shows integrated intensity of different peaks in ML, as a function of the gate voltage.

For BL (Fig. 2(b)), with application of high negative gate voltages, a peak emerges that is attributed to $X^{+}$ (also indicated in Fig. 2(a)). Fig. 2(d) shows the PL intensity as a function of gate voltage in BL WSe$_2$. $X^{0}$, $X^{-}$ and $XX^-$ show a similar doping dependence as in the ML. From the color plot of Fig. 2(b) we also observe a broad low energy indirect exciton $X^{I}$ peak $\approx$ 1.55 eV. Specific domes of high PL intensity of $X^{I}$ peaks of BL WSe$_2$ emerge as electrostatic doping is modified, which could possibly be related to resonant exciton-phonon scattering \cite{altaiary2022electrically}. The domes of high PL have not been studied in a second device, but are hinted by previous work \cite{lindlau2018role,scuri2020electrically}. A systematic experimental and theoretical study is required to understand its origin and behaviour with doping. More details are discussed in SI.
\begin{table*}[ht]
\caption{Identification of excitonic peaks in WSe$_2$ system}
\begin{tabular}{|p{13mm}|p{40mm}|p{90mm}|} 
 \hline
 Excitonic peak  & Peak position (eV) (our work) & Peak position (eV) (literature) \\
 \hline
$X^0$   & 1.736 (ML), 1.735 (tBL) & 1.733 (ML) \cite{li2019momentum}, 1.708 (ML) \cite{chen2018coulomb},  1.74 (ML) \cite{li2018revealing}  \\
\hline
$X^-$  & 1.702 (ML) $\Delta$=34 meV \newline 1.703 (tBL) $\Delta$=32 meV & 1.697 (ML)$\Delta$=36 meV \cite{li2019momentum},  1.673  (ML)$\Delta$=35 meV \cite{chen2018coulomb}, \newline 1.705 (ML)$\Delta$=35 meV \cite{li2018revealing} \\
\hline$XX$   & 1.715 (tBL) $\Delta$=20 meV  &  1.69 (ML)$\Delta$=18 meV  \cite{chen2018coulomb}, 1.715 (ML)$\Delta$=18 meV \cite{li2019momentum}, \newline 1.723eV (ML)$\Delta$=17 meV \cite{li2018revealing} \\
\hline
$XX^-$  &  1.685 (ML) $\Delta$=51 meV \newline 1.686 (tBL) $\Delta$=49 meV  & 1.684 (ML)$\Delta$=49 \cite{li2019momentum}, 1.659 (ML)$\Delta$=49 \cite{chen2018coulomb}, \newline 1.691 (ML)$\Delta$=49 \cite{li2018revealing}\\
\hline
$X^I$ (tBL) & 1.675 (20$^{\circ}$ tBL) & 1.59 (21$^{\circ}$ tBL) \cite{merkl2020twist}, 1.607 (17$^{\circ}$ tBL) 1.525, 1.512 (0$^{\circ}$ tBL) \cite{scuri2020electrically, merkl2020twist}, 1.618 (high angle, tBL) \cite{wang2018electrical}\\
\hline
$X^{I}_{p}$ (tBL) & 1.660, 1.647  & 1.5-1.7eV \cite{altaiary2021electric}  \\
\hline
$X^I$ (BL)  & 1.57 & 1.568 \cite{scuri2020electrically},   1.569 \cite{wang2018electrical}\\
\hline
\end{tabular}

\end{table*}

Fig. 3(a) shows the doping dependent energy mapping of the tBL WSe$_2$. Fig. 3(b) shows the peak intensity of different excitonic peaks in tBL as a function of doping. Multi-peak Lorentzian fits to the PL spectra for all devices at p-doping and n-doping conditions are given in SI, section II. Similar to the case of ML and BL, the $XX^{-}$ ($X^{--}$) shows an increase in intensity with electron doping, trion intensity decreases, while neutral exciton vanishes. PL peak positions as a function of back-gate voltage are discussed in SI, section III.  

Interestingly, we observe a distinct peak around 1.71 eV, which lies 20 meV below $X^{0}$. This coincides with energy spacing of $XX$ in hBN encapsulated ML, and is thus attributed similarly, with supporting arguments below \cite{li2018revealing, barbone2018charge,steinhoff2018biexciton,hao2017neutral}. Unlike $X^{-}$ and $XX^{-}$, which extends to n-doped or p-doped region, the $XX$ peak is restricted to the intrinsic region only. 
The BEs (${\Delta}$) of the neutral and charged biexciton are given as \cite{chen2018coulomb}  ${\Delta_{XX} = \hbar\omega_{X^{0}} -  \hbar\omega_{XX}}$ and ${\Delta_{XX^-} = \hbar\omega_{X^{-}} - \hbar\omega_{XX^-}}$ respectively. Here we consider that the emission is due to the dissociation of $XX^{-}$ into $X^{D}$ and $X^{-}$, and $XX$ into $X^{D}$ and $X^{0}$. Schematic of the process is given in SI, section IV. Thus, the binding energies of $XX^{-}$ and $XX$ need to be counted from the $X^{-}$ and $X^{0}$ emission energy, respectively. BEs measured thus are 19 meV for $XX$ and 17 meV for $XX^{-}$, in good agreement with theoretical predictions \cite{kylanpaa2015binding}. The energetic peak position and the doping dependence suggests that the peak at 1.71 eV is indeed the neutral biexciton peak. We have calculated similar BE from thermal activation studies (last section), thus providing further evidence for the peak assignment. We also studied the dependence of laser power on PL spectra. The $XX$ peak intensity shows a super-linear dependence on laser power, with the power exponent ($\alpha$) 1.25, while  $X$ has a power coefficient of 1. Details have been discussed in SI, section IV.

\begin{figure*}[t]
\includegraphics[width=1\linewidth]{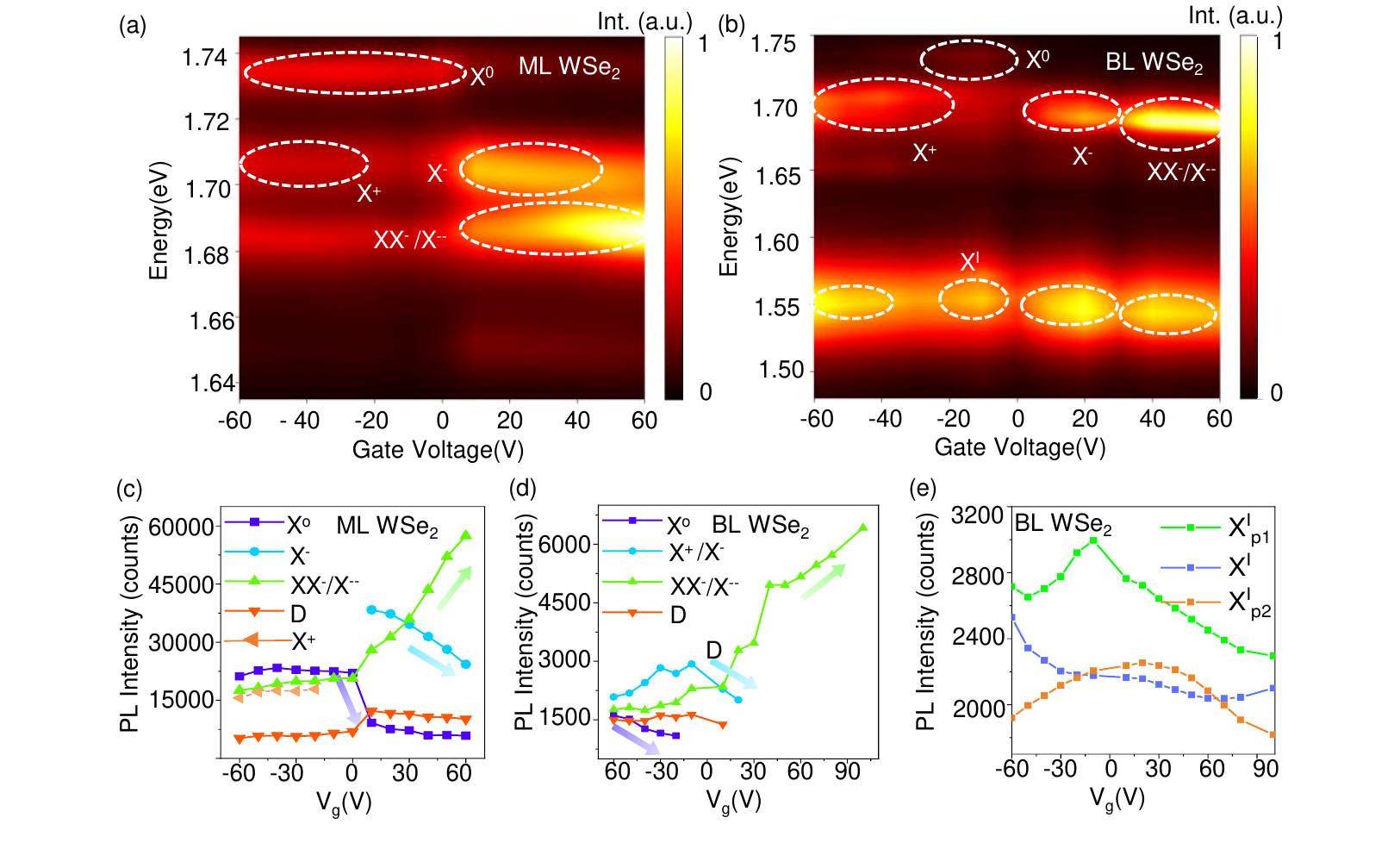}
\caption{\textbf{PL spectra of ML and BL WSe$_2$ under charge doping.} (a) and (b) show the color plot of PL spectra as a function of gate voltage in BL and ML. Regions of existence of different excitonic states under charge doping are indicated by dotted curves. (c) and (d) PL intensity of different excitonic complexes as a function of gate voltage in ML and BL respectively. (e) PL intensities of the IL excitons and phonon replicas, as a function of gate voltage in BL.}
\end{figure*}

Interestingly, $XX$ is not visible in non-encapsulated ML or BL WSe$_2$, and its observation in non-encapsulated tBL indicates the uniqueness of intermediate misorientation tBL regarding coupling and screening. Owing to the small interlayer coupling (See SI Section V for details) at large angles, the two layers of WSe$_2$ behave as a moderately decoupled system, resulting in a partial screening of the top layer from the disorder-inducing potential fluctuations of bottom SiO$_2$. This may lead to sharper spectral transitions in the tBL (where we observe the neutral XX and well separated indirect excitons) with less broadening in the peak width as compared to ML and BL, with the additional knob of valley band-alignment.

\begin{figure*}[t]
\includegraphics[width=1\linewidth]{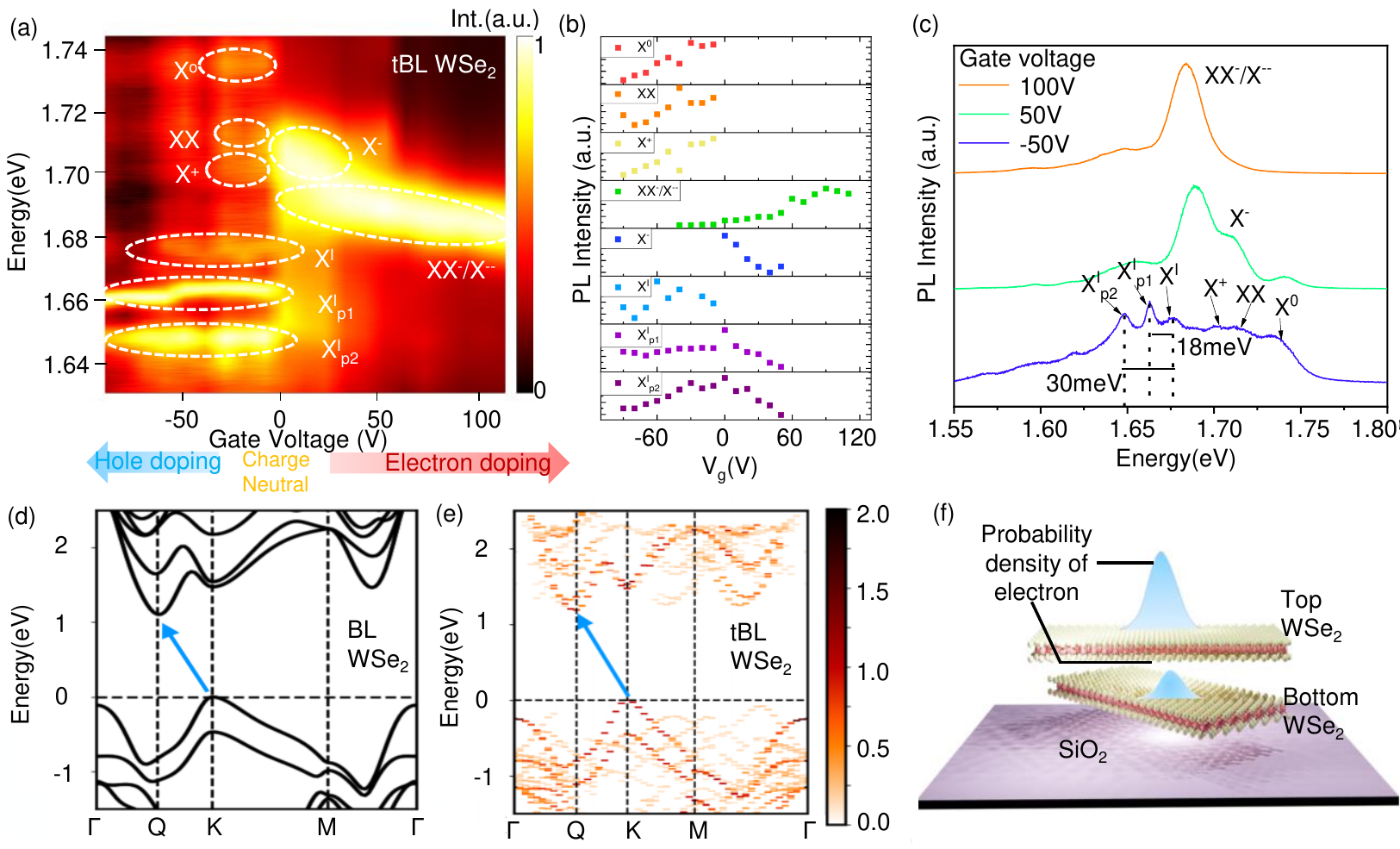}
\caption{\textbf{Electrical tuning of excitonic states in tBL WSe$_2$.} (a) Color map of PL spectra as function of gate voltage in tBL WSe$_2$. Regions of existence of different excitonic states under charge doping are indicated by dotted curves. (b) PL intensity of different excitonic species as a function of gate voltage. (c) PL spectra at 7 K at different doping densities (gate voltages). (d) The calculated band structure of BL WSe$_2$ and (e) unfolded band structure of 21$^{\circ}$ commensurate tBL WSe$_2$. The arrows show the indirect energy band gap. Note: Valence band maximum is set to zero. (f) Schematic of electron probability density distribution in tBL WSe$_2$.}
\end{figure*}

\textit{\textbf{Indirect exciton and phonon replica:}} 
We now focus on the low energy (1.5 - 1.65 eV) peaks in tBL-WSe$_2$, corresponding to the momentum indirect excitons ($X^{I}$). They are dependent on the interlayer coupling, which is significant even at intermediate twist angle of 20$^{\circ}$ \cite{scuri2020electrically,wang2018electrical}. Fig. 3(a) shows the energy mapping of the indirect excitons of tBL WSe$_2$. We note that indirect peak of the tBL WSe$_2$ is blue-shifted by 105 meV, as compared to the BL. We performed DFT calculations, as solving the Bethe-Salpeter equation for a twisted bilayer is computationally prohibitive. As the polarizability is similar in tBL and BL, we analyse the excitons qualitatively with the input from the electronic structure calculations. Since the relaxation effect is small and the moir\'{e} potential is weak in this system, the wavefunction at band edges do not localize at a particular stacking. The shift in the exciton spectra can therefore be related to the modification of the band structure induced by twist in a in 21$^{\circ}$ commensurate tBL WSe$_2$. At a large twist angle, the hybridization of the individual layers quenches significantly, resulting in reduction of the interlayer coupling, which modifies the Q-valley of the Brillouin Zone. Unfolding the band structure of 21$^\circ$ tBL to the unit cell Brillouin zone shows that the conduction band minimum has shifted from Q point to a nearby point. We call this point $Q^\prime$, which is not along $\Gamma$-K direction in unit cell Brillouin zone (see SI). The blue shift of the indirect exciton can be compared to the shift of the Q-valley in the calculated electronic band structure. The lowest indirect transition energy of tBL increases by 78 meV compared to that of BL (Fig. 3(d, e)). This compares well with experimentally observed blue-shift of the indirect peak.
\begin{figure*}
\includegraphics[width=1\linewidth]{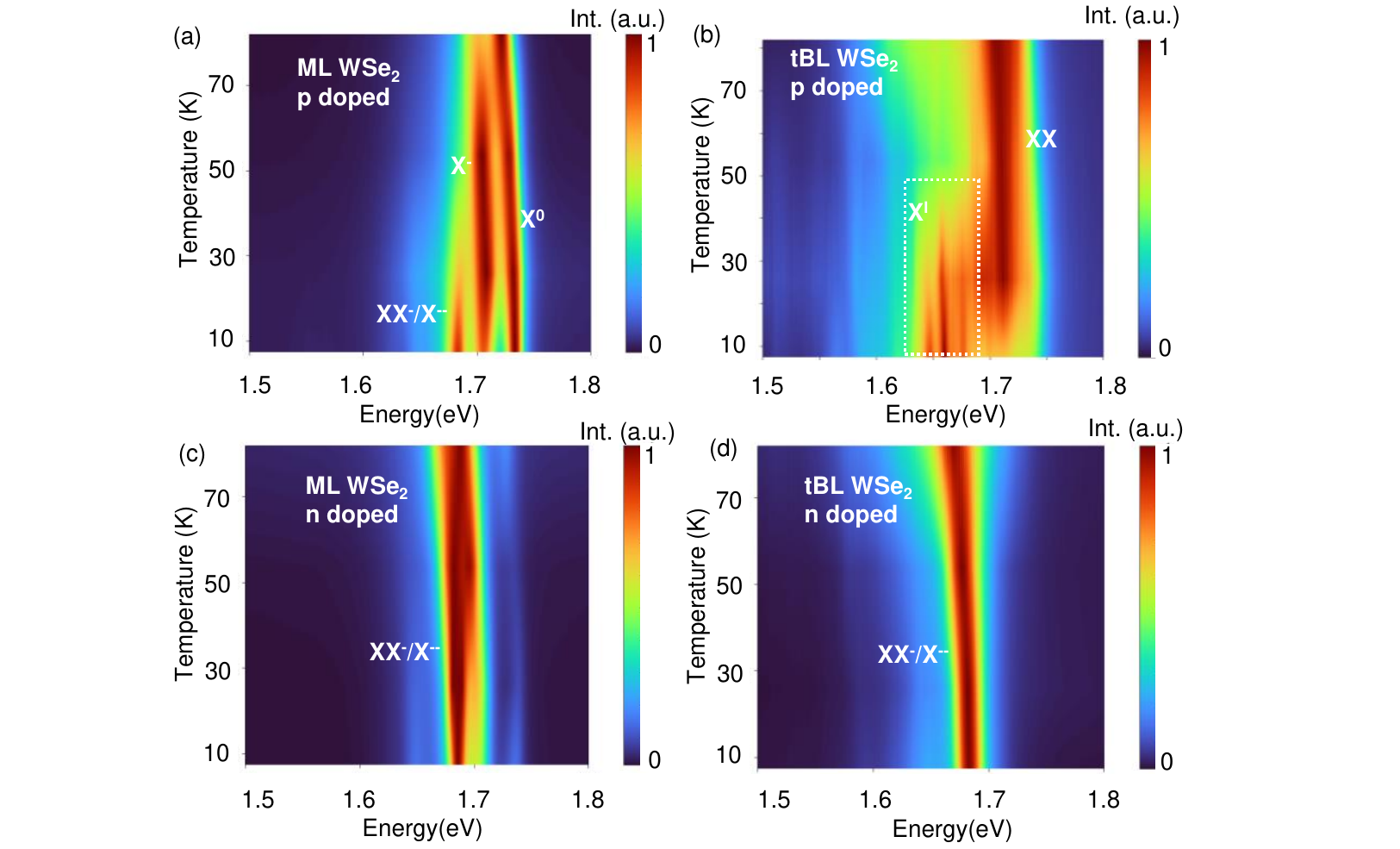}
\caption{\textbf{Thermally activated PL spectra of ML and tBL WSe$_2$.} (a) and (b) Color plot of PL spectra as a function of temperature in ML and tBL respectively, at V$_g$ = -60 V in the hole doped regime. (c) and (d) Color plot of PL spectra as a function of temperature in ML and tBL respectively, at V$_g$ = 60 V in the electron doped regime.}
\end{figure*}

Fig 3(a) and (c) show two more sharp peaks with red-shifted energies of 18 meV and 30 meV, from the $X^{I}$ emission peak. These peaks have recently been observed as two-phonon replicas ($X^{I}_{p1}$, $X^{I}_{p2}$ of exciton), where nearly resonant exciton-phonon scattering processes take place \cite{altaiary2021electric, altaiary2022electrically}. In the BL, the shifts are 20 meV and 41 meV. Additionally, we find that the phonon replicas are much brighter than their primary emission peak. In tBL, these peaks are also much sharper, suggesting a layer-dependent asymmetric electron probability density, that depends upon the twist angle of a tBL \cite{merkl2020twist, sung2020broken, weston2020atomic} (see Fig. 3(f)). More details about this are given in SI, section III.

Remarkably, the indirect excitonic states of tBL WSe$_2$ show an asymmetric dependence on doping density (see Fig. 3(c)). With increasing electron carrier density, indirect excitons strongly quench in the tBL (also see Fig. 3(a)). On the other hand, indirect excitons in BL show an oscillatory intensity profile on both the electron and hole doped side. This distinct gate dependence can be understood qualitatively by the $Q^\prime$-valley population of tBL and BL WSe$_2$. The spectral emission depends on both the oscillator strength and the electron population density. Band structure of tBL WSe$_2$ in the moir\'{e} Brillouin zone is given in SI, section V. For tBL, the energy difference ($\Delta$) between the conduction band minimum at  $Q^\prime$ (Q for BL) and the conduction band edge at K-valley is 295 meV, whereas in BL it is 366 meV (see Fig. 3(d, e)). Further, tuning the gate voltage in n-doped regime leads to an upward shift of the quasi-Fermi level of the system (See S13 in SI for details) resulting in an effective density tuning of both Q and K-valley, as well as phase filling of primarily the Q-valley. When the energy difference of the conduction band edges at the Q and K-valley reduces, the probability of photo-excited electrons relaxing to K-valley, and subsequent direct radiative transitions will be enhanced. Since momentum-direct K-K transitions are more probable than the momentum-indirect $Q-K$ transitions, negatively charged trions and biexcitons increase with n-doping and indirect excitons quickly vanish. In contrast, for the case of BL, the energy difference between Q and K-valley is much higher, and thus even in high n-doped regime, photo-excited electrons are mostly localized at the Q- valley leading to indirect exciton PL emission. Other two peaks ($X^{I}_{p1}$, $X^{I}_{p2}$) have similar doping dependence as $X^{I}$, and are thus again attributed to two-phonon replicas of $X^{I}$ \cite{altaiary2021electric}. 

\begin{figure*}[t]
\includegraphics[width=1\linewidth]{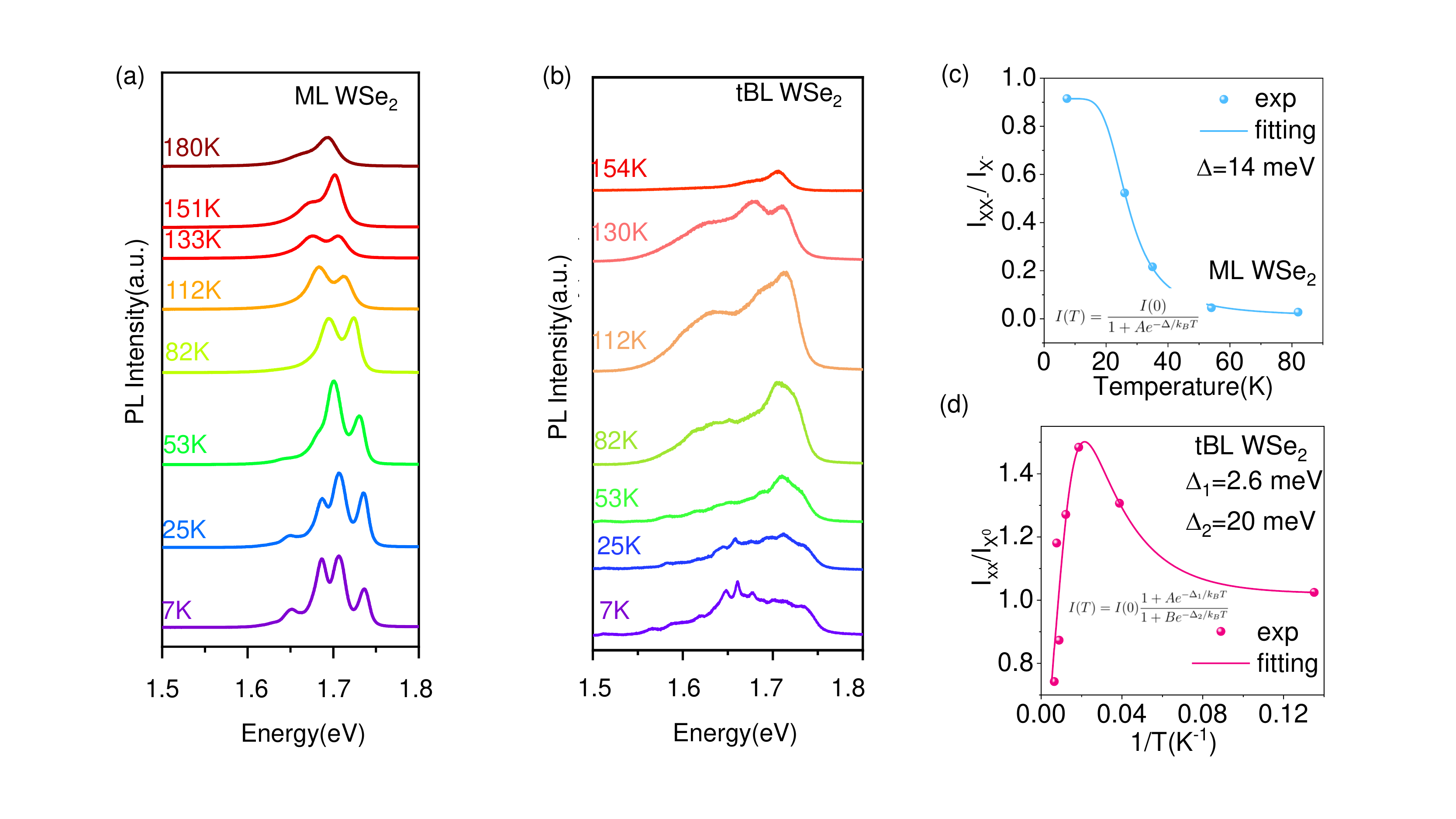}
\caption{\textbf{Temperature dependence of spectral emissions in ML and tBL WSe$_2$.} (a) and (b) PL spectra at different temperatures in ML and tBL respectively. PL spectra in (a) and (b) have been vertically shifted for clarity. The spectra has not been normalized, to aid visualization of temperature dependence of PL intensity.  (c) Normalized excitonic intensity of $XX^-$ to $X^{-}$ in ML as a function of temperature. Solid line is the fit to Eqn. (1). (d) Normalized excitonic intensity of $XX$ to $X^{0}$ in tBL as a function of temperature. Solid line is the fit to Eqn. (2).}
\end{figure*}

\textit{\textbf{Thermal disassociation of many-body charge complexes:}} To further clarify the underlying exciton interactions, we perform thermal activation measurements. Fig. 4 (a) and (b) shows the temperature dependence of ML and tBL WSe$_2$ at the p-doped region (V$g$ = -60 V), and similarly, Fig. 4(c) and (d) corresponds to the n-doped region (V$g$ = 60 V). Similar color plots for BL are given in SI, section IV. In the p-doped regime, the high temperature PL spectrum of ML WSe$_2$ is mainly dominated by $X^{0}$ due to its large binding energy of $\approx$ 400 meV. The n-doped region for ML and tBL is solely dominated by the $XX^{-}$ (or $X^{--}$) which survives up to 80 K, implying the thermal robustness of the charged complex. In the near charge-neutral region in tBL WSe$_2$, $XX$ is found to survive to much higher temperatures. At low temperatures, the optical response shifts toward the indirect PL peaks (at p-doping), along with two-phonon replica peaks. The indirect peaks are highly sensitive to temperature because the exciton distribution becomes very narrow in momentum-space at low temperature.

In Fig. 5(a) and (b) we plot the PL spectra of ML and tBL WSe$_2$ in the near charge-neutral region at different temperatures. As the temperature increases, we observe that the overall emission intensities gradually increase, and then surprisingly decrease beyond 112 K. Non-radiative recombination is a dominant process at high temperatures and leads to thermal quenching. For example, the PL intensity of MoS$_2$, decreases monotonically with increasing temperature \cite{pei2015exciton}. For WSe$_2$, on the other hand, PL intensity generally increases with temperature due to quenching of dark excitons and enhancement of the population of bright excitons. Here, we observe negative thermal quenching in the tBL, which can be understood as the thermal activation to the exciton and trion state from the low energy states, resulting in an increase in the population of the higher energy excitonic complexes \cite{shibata1998negative}. 

Analysis of the thermal activated behaviour provides insight into the binding energies, and the formation and dissociation dynamics of the excitonic complexes $XX$ and $XX^{-}$. If we consider high density of dark excitons at low temperatures, then the formation process of excitonic complexes is limited by the population of minority species, i.e. formation of $XX$ is limited by population of $X^{0}$, and $XX^{-}$ by $X^{-}$ \cite{li2018revealing}. To quantitatively understand the thermal dissociation process, we normalize the intensity of $XX^{-}$ to the intensity of $X^{-}$ in the case of ML WSe$_2$ (see Fig. 5(c)), and intensity of the $XX$ to the intensity of $X^0$ in the case of tBL WSe$_2$ (see Fig. 5(d)) \cite{chen2018coulomb}. We observe normal thermal quenching for $XX^{-}$ in the ML, and this thermally activated disassociation can be captured by using the thermal activation equation considering only one binding energy ${\Delta}$ \cite{shibata1998negative, ross2013electrical}, where $I(0)$ is the intensity at 0 K, ${\Delta}$ is the binding energy, $k_{B}$ is the Boltzmann constant, and $A$ is a fitting parameter. Using Eqn. (1) to fit our experimental data, we find that  ${\Delta}$ of $XX^{-}$ is 14 ${\pm}$ 3 meV. In our work, we have followed previously used framework (ref~\cite{chen2018coulomb}) to estimate the binding energies of biexcitons. This method gives us a rough estimate of the binding energy, since the entropy of ionization is not included (ref~\cite{kaindl2009transient,mock1978entropy}).

To understand negative thermal quenching in tBL WSe$_2$, the multi-level model for the temperature dependence of the PL peak intensities is given by the activated energy behavior \cite{shibata1998negative, huang2016probing} as shown in Fig 5.

where $A$ and $B$ are fitting parameters. $\Delta_{1}$ describes the thermal activation energy corresponding to the increment of PL intensity with temperature, whereas $\Delta_{2}$ represents the activation energy for the thermal quenching process. For a two-level system, the binding energy of $XX$ is given by $\Delta_{2}-\Delta_{1}$ = 17$\pm$5 meV. The binding energy values for both $XX$ and $XX^{-}$ are in reasonable agreement with earlier calculations ${\Delta_{XX}}$ = 20.7 meV and  ${\Delta_{XX^-}}$ = 14.9 meV \cite{kylanpaa2015binding}.

\section{Conclusion}
In summary, we have observed many-body correlated excitonic states  in ML, BL and tBL WSe$_2$. Low temperature PL reveals several higher-order excitonic states, including the neutral biexciton, which was observed only in the tBL WSe$_2$ showing the versatility of tBL in providing an interesting test-bed for light-matter interactions. In the tBL, $XX$ is dominant in the p-doped regime at high temperatures, while $XX^{-}$ is dominant in n-doped regime. Stability of these higher-order excitonic complexes at high temperatures make the tBL an exciting prototype for exploring multi-body correlated physics even at high temperatures. Remarkably, the indirect excitons in tBL vanish in the electron-doped regime due to twist-angle related modification of band-structure. Through DFT calculations, we find that  $Q^\prime$-K band-alignment is related to vanishing of indirect excitons in electron-doped regime, as well as blue shift of indirect excitons in tBL (compared to BL). In general, this work provides an improved physical understanding of the complex excitonic states through the interplay of charge, temperature and twist-angle. The studies can be extended to manipulation of excitons for applications such as exciton lasing by coupling to photonic cavity and also for harvesting high-order excitonic states for potential applications in sensing, imaging, nanophotonics and valleytronics.

\section{Materials and Methods}
\subsection{Experiment}

\textbf{PL spectroscopy:} PL spectra of WSe$_2$ flakes were taken using using Horiba LabRAM HR spectrometer. We performed the measurements at the same experimental parameters for all samples to remove any ambiguity. We used 532 nm laser with laser power 72 $\mu$W, a 50X long working distance objective with NA 0.5, an Integration time of 30 sec, and a laser spot size $\sim1.2\mu$m. Low temperature Montana cryostat was used to measure PL spectra at low temperatures.

\textbf{SHG measurements:} Polarization-dependent Second Harmonic Generation (PSHG) measurements were performed using a nonlinear microscopy imaging setup with a linearly polarized femtosecond laser (Fidelity HP-10 laser, 80 MHz repetition rate, wavelength at 1040 nm and pulse duration of 140 fs) used as an excitation source. The input laser is focused using a 20X/0.75NA objective lens, and the same objective is used to collect the backward SHG signal. The SHG signal is detected using a photomultiplier tube with a dichroic filter, a set of bandpass (520/15 nm) and short pass (890 nm) filters and a polarizer(analyzer) mounted in front of it for rejecting the fundamental source and minimizing the background noise. For the twist angle measurement, the sample is rotated from 0-90$^\circ$ in the step of 5$^{\circ}$ with respect to the laboratory horizontal axis keeping the input polarizer and the output analyzer fixed and in parallel with the laboratory horizontal axis. For each angle, SHG images of the twisted sample are acquired and later analyzed to calculate the twist angle.

\textbf{DFT Calculations:} Quantum Espresso package \cite{giannozzi2009quantum} was used to perform the density functional theory calculations of the BL and tBL WSe$_2$. The exchange correlation was approximated with local gradient approximation \cite{perdew1981self}. The unit cell Brillouin zone was sampled with 12x12 Monkhorst-Pack k-grid \cite{monkhorst1976special} and themoire Brillouin zone was scaled accordingly. The band structure  of the 21$^{\circ}$ commensurate tBL WSe$_2$ was calculated in moir\'{e} Brillouin zone and unfolded to unit cell Brillouin zone \cite{popescu2012extracting} to identify the bands edges.

\section{Acknowledgements}
The authors acknowledge financial support from U.S. Army International Technology Center Pacific (ITC-PAC) and Ministry of Electronics and Information Technology, Government of India, as well as the Supercomputer Education and Research Centre (SERC) at IISc for providing computational resources. AS acknowledges funding from IISc start-up grant.
\section{Notes}
The authors declare no competing financial interests.
\maketitle
\bibliography{main}

\begin{thebibliography}{49}%
\makeatletter
\providecommand \@ifxundefined [1]{%
 \@ifx{#1\undefined}
}%
\providecommand \@ifnum [1]{%
 \ifnum #1\expandafter \@firstoftwo
 \else \expandafter \@secondoftwo
 \fi
}%
\providecommand \@ifx [1]{%
 \ifx #1\expandafter \@firstoftwo
 \else \expandafter \@secondoftwo
 \fi
}%
\providecommand \natexlab [1]{#1}%
\providecommand \enquote  [1]{``#1''}%
\providecommand \bibnamefont  [1]{#1}%
\providecommand \bibfnamefont [1]{#1}%
\providecommand \citenamefont [1]{#1}%
\providecommand \href@noop [0]{\@secondoftwo}%
\providecommand \href [0]{\begingroup \@sanitize@url \@href}%
\providecommand \@href[1]{\@@startlink{#1}\@@href}%
\providecommand \@@href[1]{\endgroup#1\@@endlink}%
\providecommand \@sanitize@url [0]{\catcode `\\12\catcode `\$12\catcode
  `\&12\catcode `\#12\catcode `\^12\catcode `\_12\catcode `\%12\relax}%
\providecommand \@@startlink[1]{}%
\providecommand \@@endlink[0]{}%
\providecommand \url  [0]{\begingroup\@sanitize@url \@url }%
\providecommand \@url [1]{\endgroup\@href {#1}{\urlprefix }}%
\providecommand \urlprefix  [0]{URL }%
\providecommand \Eprint [0]{\href }%
\providecommand \doibase [0]{http://dx.doi.org/}%
\providecommand \selectlanguage [0]{\@gobble}%
\providecommand \bibinfo  [0]{\@secondoftwo}%
\providecommand \bibfield  [0]{\@secondoftwo}%
\providecommand \translation [1]{[#1]}%
\providecommand \BibitemOpen [0]{}%
\providecommand \bibitemStop [0]{}%
\providecommand \bibitemNoStop [0]{.\EOS\space}%
\providecommand \EOS [0]{\spacefactor3000\relax}%
\providecommand \BibitemShut  [1]{\csname bibitem#1\endcsname}%
\let\auto@bib@innerbib\@empty
%</preamble>
\bibitem [{\citenamefont {Yu}\ \emph {et~al.}(2017)\citenamefont {Yu},
  \citenamefont {Liu}, \citenamefont {Tang}, \citenamefont {Xu},\ and\
  \citenamefont {Yao}}]{yu2017moire}%
  \BibitemOpen
  \bibfield  {author} {\bibinfo {author} {\bibfnamefont {H.}~\bibnamefont
  {Yu}}, \bibinfo {author} {\bibfnamefont {G.-B.}\ \bibnamefont {Liu}},
  \bibinfo {author} {\bibfnamefont {J.}~\bibnamefont {Tang}}, \bibinfo {author}
  {\bibfnamefont {X.}~\bibnamefont {Xu}}, \ and\ \bibinfo {author}
  {\bibfnamefont {W.}~\bibnamefont {Yao}},\ }\href@noop {} {\bibfield
  {journal} {\bibinfo  {journal} {Science advances}\ }\textbf {\bibinfo
  {volume} {3}},\ \bibinfo {pages} {e1701696} (\bibinfo {year}
  {2017})}\BibitemShut {NoStop}%
\bibitem [{\citenamefont {Tran}\ \emph {et~al.}(2020)\citenamefont {Tran},
  \citenamefont {Choi},\ and\ \citenamefont {Singh}}]{tran2020moire}%
  \BibitemOpen
  \bibfield  {author} {\bibinfo {author} {\bibfnamefont {K.}~\bibnamefont
  {Tran}}, \bibinfo {author} {\bibfnamefont {J.}~\bibnamefont {Choi}}, \ and\
  \bibinfo {author} {\bibfnamefont {A.}~\bibnamefont {Singh}},\ }\href@noop {}
  {\bibfield  {journal} {\bibinfo  {journal} {2D Materials}\ }\textbf {\bibinfo
  {volume} {8}},\ \bibinfo {pages} {022002} (\bibinfo {year}
  {2020})}\BibitemShut {NoStop}%
\bibitem [{\citenamefont {Andersen}\ \emph {et~al.}(2021)\citenamefont
  {Andersen}, \citenamefont {Scuri}, \citenamefont {Sushko}, \citenamefont
  {De~Greve}, \citenamefont {Sung}, \citenamefont {Zhou}, \citenamefont {Wild},
  \citenamefont {Gelly}, \citenamefont {Heo}, \citenamefont {B{\'e}rub{\'e}}
  \emph {et~al.}}]{andersenexcitons}%
  \BibitemOpen
  \bibfield  {author} {\bibinfo {author} {\bibfnamefont {T.~I.}\ \bibnamefont
  {Andersen}}, \bibinfo {author} {\bibfnamefont {G.}~\bibnamefont {Scuri}},
  \bibinfo {author} {\bibfnamefont {A.}~\bibnamefont {Sushko}}, \bibinfo
  {author} {\bibfnamefont {K.}~\bibnamefont {De~Greve}}, \bibinfo {author}
  {\bibfnamefont {J.}~\bibnamefont {Sung}}, \bibinfo {author} {\bibfnamefont
  {Y.}~\bibnamefont {Zhou}}, \bibinfo {author} {\bibfnamefont {D.~S.}\
  \bibnamefont {Wild}}, \bibinfo {author} {\bibfnamefont {R.~J.}\ \bibnamefont
  {Gelly}}, \bibinfo {author} {\bibfnamefont {H.}~\bibnamefont {Heo}}, \bibinfo
  {author} {\bibfnamefont {D.}~\bibnamefont {B{\'e}rub{\'e}}},  \emph
  {et~al.},\ }\href@noop {} {\bibfield  {journal} {\bibinfo  {journal} {Nature
  Materials}\ }\textbf {\bibinfo {volume} {20}},\ \bibinfo {pages} {1}
  (\bibinfo {year} {2021})}\BibitemShut {NoStop}%
\bibitem [{\citenamefont {Tran}\ \emph {et~al.}(2019)\citenamefont {Tran},
  \citenamefont {Moody}, \citenamefont {Wu}, \citenamefont {Lu}, \citenamefont
  {Choi}, \citenamefont {Kim}, \citenamefont {Rai}, \citenamefont {Sanchez},
  \citenamefont {Quan}, \citenamefont {Singh} \emph
  {et~al.}}]{tran2019evidence}%
  \BibitemOpen
  \bibfield  {author} {\bibinfo {author} {\bibfnamefont {K.}~\bibnamefont
  {Tran}}, \bibinfo {author} {\bibfnamefont {G.}~\bibnamefont {Moody}},
  \bibinfo {author} {\bibfnamefont {F.}~\bibnamefont {Wu}}, \bibinfo {author}
  {\bibfnamefont {X.}~\bibnamefont {Lu}}, \bibinfo {author} {\bibfnamefont
  {J.}~\bibnamefont {Choi}}, \bibinfo {author} {\bibfnamefont {K.}~\bibnamefont
  {Kim}}, \bibinfo {author} {\bibfnamefont {A.}~\bibnamefont {Rai}}, \bibinfo
  {author} {\bibfnamefont {D.~A.}\ \bibnamefont {Sanchez}}, \bibinfo {author}
  {\bibfnamefont {J.}~\bibnamefont {Quan}}, \bibinfo {author} {\bibfnamefont
  {A.}~\bibnamefont {Singh}},  \emph {et~al.},\ }\href@noop {} {\bibfield
  {journal} {\bibinfo  {journal} {Nature}\ }\textbf {\bibinfo {volume} {567}},\
  \bibinfo {pages} {71} (\bibinfo {year} {2019})}\BibitemShut {NoStop}%
\bibitem [{\citenamefont {Jin}\ \emph {et~al.}(2019)\citenamefont {Jin},
  \citenamefont {Regan}, \citenamefont {Yan}, \citenamefont {Utama},
  \citenamefont {Wang}, \citenamefont {Zhao}, \citenamefont {Qin},
  \citenamefont {Yang}, \citenamefont {Zheng}, \citenamefont {Shi} \emph
  {et~al.}}]{jin2019observation}%
  \BibitemOpen
  \bibfield  {author} {\bibinfo {author} {\bibfnamefont {C.}~\bibnamefont
  {Jin}}, \bibinfo {author} {\bibfnamefont {E.~C.}\ \bibnamefont {Regan}},
  \bibinfo {author} {\bibfnamefont {A.}~\bibnamefont {Yan}}, \bibinfo {author}
  {\bibfnamefont {M.~I.~B.}\ \bibnamefont {Utama}}, \bibinfo {author}
  {\bibfnamefont {D.}~\bibnamefont {Wang}}, \bibinfo {author} {\bibfnamefont
  {S.}~\bibnamefont {Zhao}}, \bibinfo {author} {\bibfnamefont {Y.}~\bibnamefont
  {Qin}}, \bibinfo {author} {\bibfnamefont {S.}~\bibnamefont {Yang}}, \bibinfo
  {author} {\bibfnamefont {Z.}~\bibnamefont {Zheng}}, \bibinfo {author}
  {\bibfnamefont {S.}~\bibnamefont {Shi}},  \emph {et~al.},\ }\href@noop {}
  {\bibfield  {journal} {\bibinfo  {journal} {Nature}\ }\textbf {\bibinfo
  {volume} {567}},\ \bibinfo {pages} {76} (\bibinfo {year} {2019})}\BibitemShut
  {NoStop}%
\bibitem [{\citenamefont {Alexeev}\ \emph {et~al.}(2019)\citenamefont
  {Alexeev}, \citenamefont {Ruiz-Tijerina}, \citenamefont {Danovich},
  \citenamefont {Hamer}, \citenamefont {Terry}, \citenamefont {Nayak},
  \citenamefont {Ahn}, \citenamefont {Pak}, \citenamefont {Lee}, \citenamefont
  {Sohn} \emph {et~al.}}]{alexeev2019resonantly}%
  \BibitemOpen
  \bibfield  {author} {\bibinfo {author} {\bibfnamefont {E.~M.}\ \bibnamefont
  {Alexeev}}, \bibinfo {author} {\bibfnamefont {D.~A.}\ \bibnamefont
  {Ruiz-Tijerina}}, \bibinfo {author} {\bibfnamefont {M.}~\bibnamefont
  {Danovich}}, \bibinfo {author} {\bibfnamefont {M.~J.}\ \bibnamefont {Hamer}},
  \bibinfo {author} {\bibfnamefont {D.~J.}\ \bibnamefont {Terry}}, \bibinfo
  {author} {\bibfnamefont {P.~K.}\ \bibnamefont {Nayak}}, \bibinfo {author}
  {\bibfnamefont {S.}~\bibnamefont {Ahn}}, \bibinfo {author} {\bibfnamefont
  {S.}~\bibnamefont {Pak}}, \bibinfo {author} {\bibfnamefont {J.}~\bibnamefont
  {Lee}}, \bibinfo {author} {\bibfnamefont {J.~I.}\ \bibnamefont {Sohn}},
  \emph {et~al.},\ }\href@noop {} {\bibfield  {journal} {\bibinfo  {journal}
  {Nature}\ }\textbf {\bibinfo {volume} {567}},\ \bibinfo {pages} {81}
  (\bibinfo {year} {2019})}\BibitemShut {NoStop}%
\bibitem [{\citenamefont {Baek}\ \emph {et~al.}(2020)\citenamefont {Baek},
  \citenamefont {Brotons-Gisbert}, \citenamefont {Koong}, \citenamefont
  {Campbell}, \citenamefont {Rambach}, \citenamefont {Watanabe}, \citenamefont
  {Taniguchi},\ and\ \citenamefont {Gerardot}}]{baek2020highly}%
  \BibitemOpen
  \bibfield  {author} {\bibinfo {author} {\bibfnamefont {H.}~\bibnamefont
  {Baek}}, \bibinfo {author} {\bibfnamefont {M.}~\bibnamefont
  {Brotons-Gisbert}}, \bibinfo {author} {\bibfnamefont {Z.}~\bibnamefont
  {Koong}}, \bibinfo {author} {\bibfnamefont {A.}~\bibnamefont {Campbell}},
  \bibinfo {author} {\bibfnamefont {M.}~\bibnamefont {Rambach}}, \bibinfo
  {author} {\bibfnamefont {K.}~\bibnamefont {Watanabe}}, \bibinfo {author}
  {\bibfnamefont {T.}~\bibnamefont {Taniguchi}}, \ and\ \bibinfo {author}
  {\bibfnamefont {B.~D.}\ \bibnamefont {Gerardot}},\ }\href@noop {} {\bibfield
  {journal} {\bibinfo  {journal} {Science advances}\ }\textbf {\bibinfo
  {volume} {6}},\ \bibinfo {pages} {eaba8526} (\bibinfo {year}
  {2020})}\BibitemShut {NoStop}%
\bibitem [{\citenamefont {Brotons-Gisbert}\ \emph {et~al.}(2021)\citenamefont
  {Brotons-Gisbert}, \citenamefont {Baek}, \citenamefont {Campbell},
  \citenamefont {Watanabe}, \citenamefont {Taniguchi},\ and\ \citenamefont
  {Gerardot}}]{brotons2021moir}%
  \BibitemOpen
  \bibfield  {author} {\bibinfo {author} {\bibfnamefont {M.}~\bibnamefont
  {Brotons-Gisbert}}, \bibinfo {author} {\bibfnamefont {H.}~\bibnamefont
  {Baek}}, \bibinfo {author} {\bibfnamefont {A.}~\bibnamefont {Campbell}},
  \bibinfo {author} {\bibfnamefont {K.}~\bibnamefont {Watanabe}}, \bibinfo
  {author} {\bibfnamefont {T.}~\bibnamefont {Taniguchi}}, \ and\ \bibinfo
  {author} {\bibfnamefont {B.~D.}\ \bibnamefont {Gerardot}},\ }\href@noop {}
  {\bibfield  {journal} {\bibinfo  {journal} {arXiv preprint arXiv:2101.07747}\
  } (\bibinfo {year} {2021})}\BibitemShut {NoStop}%
\bibitem [{\citenamefont {Tang}\ \emph {et~al.}(2020)\citenamefont {Tang},
  \citenamefont {Gu}, \citenamefont {Liu}, \citenamefont {Watanabe},
  \citenamefont {Taniguchi}, \citenamefont {Hone}, \citenamefont {Mak},\ and\
  \citenamefont {Shan}}]{tang2020tuning}%
  \BibitemOpen
  \bibfield  {author} {\bibinfo {author} {\bibfnamefont {Y.}~\bibnamefont
  {Tang}}, \bibinfo {author} {\bibfnamefont {J.}~\bibnamefont {Gu}}, \bibinfo
  {author} {\bibfnamefont {S.}~\bibnamefont {Liu}}, \bibinfo {author}
  {\bibfnamefont {K.}~\bibnamefont {Watanabe}}, \bibinfo {author}
  {\bibfnamefont {T.}~\bibnamefont {Taniguchi}}, \bibinfo {author}
  {\bibfnamefont {J.}~\bibnamefont {Hone}}, \bibinfo {author} {\bibfnamefont
  {K.~F.}\ \bibnamefont {Mak}}, \ and\ \bibinfo {author} {\bibfnamefont
  {J.}~\bibnamefont {Shan}},\ }\href@noop {} {\bibfield  {journal} {\bibinfo
  {journal} {Nature Nanotechnology}\ }\textbf {\bibinfo {volume} {16}},\
  \bibinfo {pages} {1} (\bibinfo {year} {2020})}\BibitemShut {NoStop}%
\bibitem [{\citenamefont {Regan}\ \emph {et~al.}(2020)\citenamefont {Regan},
  \citenamefont {Wang}, \citenamefont {Jin}, \citenamefont {Utama},
  \citenamefont {Gao}, \citenamefont {Wei}, \citenamefont {Zhao}, \citenamefont
  {Zhao}, \citenamefont {Zhang}, \citenamefont {Yumigeta} \emph
  {et~al.}}]{regan2020mott}%
  \BibitemOpen
  \bibfield  {author} {\bibinfo {author} {\bibfnamefont {E.~C.}\ \bibnamefont
  {Regan}}, \bibinfo {author} {\bibfnamefont {D.}~\bibnamefont {Wang}},
  \bibinfo {author} {\bibfnamefont {C.}~\bibnamefont {Jin}}, \bibinfo {author}
  {\bibfnamefont {M.~I.~B.}\ \bibnamefont {Utama}}, \bibinfo {author}
  {\bibfnamefont {B.}~\bibnamefont {Gao}}, \bibinfo {author} {\bibfnamefont
  {X.}~\bibnamefont {Wei}}, \bibinfo {author} {\bibfnamefont {S.}~\bibnamefont
  {Zhao}}, \bibinfo {author} {\bibfnamefont {W.}~\bibnamefont {Zhao}}, \bibinfo
  {author} {\bibfnamefont {Z.}~\bibnamefont {Zhang}}, \bibinfo {author}
  {\bibfnamefont {K.}~\bibnamefont {Yumigeta}},  \emph {et~al.},\ }\href@noop
  {} {\bibfield  {journal} {\bibinfo  {journal} {Nature}\ }\textbf {\bibinfo
  {volume} {579}},\ \bibinfo {pages} {359} (\bibinfo {year}
  {2020})}\BibitemShut {NoStop}%
\bibitem [{\citenamefont {Jin}\ \emph {et~al.}(2020)\citenamefont {Jin},
  \citenamefont {Tao}, \citenamefont {Li}, \citenamefont {Xu}, \citenamefont
  {Tang}, \citenamefont {Zhu}, \citenamefont {Liu}, \citenamefont {Watanabe},
  \citenamefont {Taniguchi}, \citenamefont {Hone} \emph
  {et~al.}}]{jin2020stripe}%
  \BibitemOpen
  \bibfield  {author} {\bibinfo {author} {\bibfnamefont {C.}~\bibnamefont
  {Jin}}, \bibinfo {author} {\bibfnamefont {Z.}~\bibnamefont {Tao}}, \bibinfo
  {author} {\bibfnamefont {T.}~\bibnamefont {Li}}, \bibinfo {author}
  {\bibfnamefont {Y.}~\bibnamefont {Xu}}, \bibinfo {author} {\bibfnamefont
  {Y.}~\bibnamefont {Tang}}, \bibinfo {author} {\bibfnamefont {J.}~\bibnamefont
  {Zhu}}, \bibinfo {author} {\bibfnamefont {S.}~\bibnamefont {Liu}}, \bibinfo
  {author} {\bibfnamefont {K.}~\bibnamefont {Watanabe}}, \bibinfo {author}
  {\bibfnamefont {T.}~\bibnamefont {Taniguchi}}, \bibinfo {author}
  {\bibfnamefont {J.~C.}\ \bibnamefont {Hone}},  \emph {et~al.},\ }\href@noop
  {} {\bibfield  {journal} {\bibinfo  {journal} {arXiv preprint
  arXiv:2007.12068}\ } (\bibinfo {year} {2020})}\BibitemShut {NoStop}%
\bibitem [{\citenamefont {You}\ \emph {et~al.}(2015)\citenamefont {You},
  \citenamefont {Zhang}, \citenamefont {Berkelbach}, \citenamefont {Hybertsen},
  \citenamefont {Reichman},\ and\ \citenamefont {Heinz}}]{you2015observation}%
  \BibitemOpen
  \bibfield  {author} {\bibinfo {author} {\bibfnamefont {Y.}~\bibnamefont
  {You}}, \bibinfo {author} {\bibfnamefont {X.-X.}\ \bibnamefont {Zhang}},
  \bibinfo {author} {\bibfnamefont {T.~C.}\ \bibnamefont {Berkelbach}},
  \bibinfo {author} {\bibfnamefont {M.~S.}\ \bibnamefont {Hybertsen}}, \bibinfo
  {author} {\bibfnamefont {D.~R.}\ \bibnamefont {Reichman}}, \ and\ \bibinfo
  {author} {\bibfnamefont {T.~F.}\ \bibnamefont {Heinz}},\ }\href@noop {}
  {\bibfield  {journal} {\bibinfo  {journal} {Nature Physics}\ }\textbf
  {\bibinfo {volume} {11}},\ \bibinfo {pages} {477} (\bibinfo {year}
  {2015})}\BibitemShut {NoStop}%
\bibitem [{\citenamefont {He}\ \emph {et~al.}(2020)\citenamefont {He},
  \citenamefont {Rivera}, \citenamefont {Van~Tuan}, \citenamefont {Wilson},
  \citenamefont {Yang}, \citenamefont {Taniguchi}, \citenamefont {Watanabe},
  \citenamefont {Yan}, \citenamefont {Mandrus}, \citenamefont {Yu} \emph
  {et~al.}}]{he2020valley}%
  \BibitemOpen
  \bibfield  {author} {\bibinfo {author} {\bibfnamefont {M.}~\bibnamefont
  {He}}, \bibinfo {author} {\bibfnamefont {P.}~\bibnamefont {Rivera}}, \bibinfo
  {author} {\bibfnamefont {D.}~\bibnamefont {Van~Tuan}}, \bibinfo {author}
  {\bibfnamefont {N.~P.}\ \bibnamefont {Wilson}}, \bibinfo {author}
  {\bibfnamefont {M.}~\bibnamefont {Yang}}, \bibinfo {author} {\bibfnamefont
  {T.}~\bibnamefont {Taniguchi}}, \bibinfo {author} {\bibfnamefont
  {K.}~\bibnamefont {Watanabe}}, \bibinfo {author} {\bibfnamefont
  {J.}~\bibnamefont {Yan}}, \bibinfo {author} {\bibfnamefont {D.~G.}\
  \bibnamefont {Mandrus}}, \bibinfo {author} {\bibfnamefont {H.}~\bibnamefont
  {Yu}},  \emph {et~al.},\ }\href@noop {} {\bibfield  {journal} {\bibinfo
  {journal} {Nature communications}\ }\textbf {\bibinfo {volume} {11}},\
  \bibinfo {pages} {1} (\bibinfo {year} {2020})}\BibitemShut {NoStop}%
\bibitem [{\citenamefont {Latini}\ \emph {et~al.}(2017)\citenamefont {Latini},
  \citenamefont {Winther}, \citenamefont {Olsen},\ and\ \citenamefont
  {Thygesen}}]{latini2017interlayer}%
  \BibitemOpen
  \bibfield  {author} {\bibinfo {author} {\bibfnamefont {S.}~\bibnamefont
  {Latini}}, \bibinfo {author} {\bibfnamefont {K.~T.}\ \bibnamefont {Winther}},
  \bibinfo {author} {\bibfnamefont {T.}~\bibnamefont {Olsen}}, \ and\ \bibinfo
  {author} {\bibfnamefont {K.~S.}\ \bibnamefont {Thygesen}},\ }\href@noop {}
  {\bibfield  {journal} {\bibinfo  {journal} {Nano letters}\ }\textbf {\bibinfo
  {volume} {17}},\ \bibinfo {pages} {938} (\bibinfo {year} {2017})}\BibitemShut
  {NoStop}%
\bibitem [{\citenamefont {Huang}\ \emph {et~al.}(2014)\citenamefont {Huang},
  \citenamefont {Ling}, \citenamefont {Liang}, \citenamefont {Kong},
  \citenamefont {Terrones}, \citenamefont {Meunier},\ and\ \citenamefont
  {Dresselhaus}}]{huang2014probing}%
  \BibitemOpen
  \bibfield  {author} {\bibinfo {author} {\bibfnamefont {S.}~\bibnamefont
  {Huang}}, \bibinfo {author} {\bibfnamefont {X.}~\bibnamefont {Ling}},
  \bibinfo {author} {\bibfnamefont {L.}~\bibnamefont {Liang}}, \bibinfo
  {author} {\bibfnamefont {J.}~\bibnamefont {Kong}}, \bibinfo {author}
  {\bibfnamefont {H.}~\bibnamefont {Terrones}}, \bibinfo {author}
  {\bibfnamefont {V.}~\bibnamefont {Meunier}}, \ and\ \bibinfo {author}
  {\bibfnamefont {M.~S.}\ \bibnamefont {Dresselhaus}},\ }\href@noop {}
  {\bibfield  {journal} {\bibinfo  {journal} {Nano letters}\ }\textbf {\bibinfo
  {volume} {14}},\ \bibinfo {pages} {5500} (\bibinfo {year}
  {2014})}\BibitemShut {NoStop}%
\bibitem [{\citenamefont {Liu}\ \emph {et~al.}(2014)\citenamefont {Liu},
  \citenamefont {Zhang}, \citenamefont {Cao}, \citenamefont {Jin},
  \citenamefont {Qiu}, \citenamefont {Zhou}, \citenamefont {Zettl},
  \citenamefont {Yang}, \citenamefont {Louie},\ and\ \citenamefont
  {Wang}}]{liu2014evolution}%
  \BibitemOpen
  \bibfield  {author} {\bibinfo {author} {\bibfnamefont {K.}~\bibnamefont
  {Liu}}, \bibinfo {author} {\bibfnamefont {L.}~\bibnamefont {Zhang}}, \bibinfo
  {author} {\bibfnamefont {T.}~\bibnamefont {Cao}}, \bibinfo {author}
  {\bibfnamefont {C.}~\bibnamefont {Jin}}, \bibinfo {author} {\bibfnamefont
  {D.}~\bibnamefont {Qiu}}, \bibinfo {author} {\bibfnamefont {Q.}~\bibnamefont
  {Zhou}}, \bibinfo {author} {\bibfnamefont {A.}~\bibnamefont {Zettl}},
  \bibinfo {author} {\bibfnamefont {P.}~\bibnamefont {Yang}}, \bibinfo {author}
  {\bibfnamefont {S.~G.}\ \bibnamefont {Louie}}, \ and\ \bibinfo {author}
  {\bibfnamefont {F.}~\bibnamefont {Wang}},\ }\href@noop {} {\bibfield
  {journal} {\bibinfo  {journal} {Nature communications}\ }\textbf {\bibinfo
  {volume} {5}},\ \bibinfo {pages} {1} (\bibinfo {year} {2014})}\BibitemShut
  {NoStop}%
\bibitem [{\citenamefont {Ross}\ \emph {et~al.}(2013)\citenamefont {Ross},
  \citenamefont {Wu}, \citenamefont {Yu}, \citenamefont {Ghimire},
  \citenamefont {Jones}, \citenamefont {Aivazian}, \citenamefont {Yan},
  \citenamefont {Mandrus}, \citenamefont {Xiao}, \citenamefont {Yao} \emph
  {et~al.}}]{ross2013electrical}%
  \BibitemOpen
  \bibfield  {author} {\bibinfo {author} {\bibfnamefont {J.~S.}\ \bibnamefont
  {Ross}}, \bibinfo {author} {\bibfnamefont {S.}~\bibnamefont {Wu}}, \bibinfo
  {author} {\bibfnamefont {H.}~\bibnamefont {Yu}}, \bibinfo {author}
  {\bibfnamefont {N.~J.}\ \bibnamefont {Ghimire}}, \bibinfo {author}
  {\bibfnamefont {A.~M.}\ \bibnamefont {Jones}}, \bibinfo {author}
  {\bibfnamefont {G.}~\bibnamefont {Aivazian}}, \bibinfo {author}
  {\bibfnamefont {J.}~\bibnamefont {Yan}}, \bibinfo {author} {\bibfnamefont
  {D.~G.}\ \bibnamefont {Mandrus}}, \bibinfo {author} {\bibfnamefont
  {D.}~\bibnamefont {Xiao}}, \bibinfo {author} {\bibfnamefont {W.}~\bibnamefont
  {Yao}},  \emph {et~al.},\ }\href@noop {} {\bibfield  {journal} {\bibinfo
  {journal} {Nature communications}\ }\textbf {\bibinfo {volume} {4}},\
  \bibinfo {pages} {1} (\bibinfo {year} {2013})}\BibitemShut {NoStop}%
\bibitem [{\citenamefont {Li}\ \emph {et~al.}(2018)\citenamefont {Li},
  \citenamefont {Wang}, \citenamefont {Lu}, \citenamefont {Jin}, \citenamefont
  {Chen}, \citenamefont {Meng}, \citenamefont {Lian}, \citenamefont
  {Taniguchi}, \citenamefont {Watanabe}, \citenamefont {Zhang} \emph
  {et~al.}}]{li2018revealing}%
  \BibitemOpen
  \bibfield  {author} {\bibinfo {author} {\bibfnamefont {Z.}~\bibnamefont
  {Li}}, \bibinfo {author} {\bibfnamefont {T.}~\bibnamefont {Wang}}, \bibinfo
  {author} {\bibfnamefont {Z.}~\bibnamefont {Lu}}, \bibinfo {author}
  {\bibfnamefont {C.}~\bibnamefont {Jin}}, \bibinfo {author} {\bibfnamefont
  {Y.}~\bibnamefont {Chen}}, \bibinfo {author} {\bibfnamefont {Y.}~\bibnamefont
  {Meng}}, \bibinfo {author} {\bibfnamefont {Z.}~\bibnamefont {Lian}}, \bibinfo
  {author} {\bibfnamefont {T.}~\bibnamefont {Taniguchi}}, \bibinfo {author}
  {\bibfnamefont {K.}~\bibnamefont {Watanabe}}, \bibinfo {author}
  {\bibfnamefont {S.}~\bibnamefont {Zhang}},  \emph {et~al.},\ }\href@noop {}
  {\bibfield  {journal} {\bibinfo  {journal} {Nature communications}\ }\textbf
  {\bibinfo {volume} {9}},\ \bibinfo {pages} {1} (\bibinfo {year}
  {2018})}\BibitemShut {NoStop}%
\bibitem [{\citenamefont {He}\ \emph {et~al.}(2014)\citenamefont {He},
  \citenamefont {Kumar}, \citenamefont {Zhao}, \citenamefont {Wang},
  \citenamefont {Mak}, \citenamefont {Zhao},\ and\ \citenamefont
  {Shan}}]{he2014tightly}%
  \BibitemOpen
  \bibfield  {author} {\bibinfo {author} {\bibfnamefont {K.}~\bibnamefont
  {He}}, \bibinfo {author} {\bibfnamefont {N.}~\bibnamefont {Kumar}}, \bibinfo
  {author} {\bibfnamefont {L.}~\bibnamefont {Zhao}}, \bibinfo {author}
  {\bibfnamefont {Z.}~\bibnamefont {Wang}}, \bibinfo {author} {\bibfnamefont
  {K.~F.}\ \bibnamefont {Mak}}, \bibinfo {author} {\bibfnamefont
  {H.}~\bibnamefont {Zhao}}, \ and\ \bibinfo {author} {\bibfnamefont
  {J.}~\bibnamefont {Shan}},\ }\href@noop {} {\bibfield  {journal} {\bibinfo
  {journal} {Physical review letters}\ }\textbf {\bibinfo {volume} {113}},\
  \bibinfo {pages} {026803} (\bibinfo {year} {2014})}\BibitemShut {NoStop}%
\bibitem [{\citenamefont {Wang}\ \emph {et~al.}(2014)\citenamefont {Wang},
  \citenamefont {Marie}, \citenamefont {Bouet}, \citenamefont {Vidal},
  \citenamefont {Balocchi}, \citenamefont {Amand}, \citenamefont {Lagarde},\
  and\ \citenamefont {Urbaszek}}]{wang2014exciton}%
  \BibitemOpen
  \bibfield  {author} {\bibinfo {author} {\bibfnamefont {G.}~\bibnamefont
  {Wang}}, \bibinfo {author} {\bibfnamefont {X.}~\bibnamefont {Marie}},
  \bibinfo {author} {\bibfnamefont {L.}~\bibnamefont {Bouet}}, \bibinfo
  {author} {\bibfnamefont {M.}~\bibnamefont {Vidal}}, \bibinfo {author}
  {\bibfnamefont {A.}~\bibnamefont {Balocchi}}, \bibinfo {author}
  {\bibfnamefont {T.}~\bibnamefont {Amand}}, \bibinfo {author} {\bibfnamefont
  {D.}~\bibnamefont {Lagarde}}, \ and\ \bibinfo {author} {\bibfnamefont
  {B.}~\bibnamefont {Urbaszek}},\ }\href@noop {} {\bibfield  {journal}
  {\bibinfo  {journal} {Applied Physics Letters}\ }\textbf {\bibinfo {volume}
  {105}},\ \bibinfo {pages} {182105} (\bibinfo {year} {2014})}\BibitemShut
  {NoStop}%
\bibitem [{\citenamefont {Lindlau}\ \emph {et~al.}(2018)\citenamefont
  {Lindlau}, \citenamefont {Selig}, \citenamefont {Neumann}, \citenamefont
  {Colombier}, \citenamefont {F{\"o}rste}, \citenamefont {Funk}, \citenamefont
  {F{\"o}rg}, \citenamefont {Kim}, \citenamefont {Bergh{\"a}user},
  \citenamefont {Taniguchi} \emph {et~al.}}]{lindlau2018role}%
  \BibitemOpen
  \bibfield  {author} {\bibinfo {author} {\bibfnamefont {J.}~\bibnamefont
  {Lindlau}}, \bibinfo {author} {\bibfnamefont {M.}~\bibnamefont {Selig}},
  \bibinfo {author} {\bibfnamefont {A.}~\bibnamefont {Neumann}}, \bibinfo
  {author} {\bibfnamefont {L.}~\bibnamefont {Colombier}}, \bibinfo {author}
  {\bibfnamefont {J.}~\bibnamefont {F{\"o}rste}}, \bibinfo {author}
  {\bibfnamefont {V.}~\bibnamefont {Funk}}, \bibinfo {author} {\bibfnamefont
  {M.}~\bibnamefont {F{\"o}rg}}, \bibinfo {author} {\bibfnamefont
  {J.}~\bibnamefont {Kim}}, \bibinfo {author} {\bibfnamefont {G.}~\bibnamefont
  {Bergh{\"a}user}}, \bibinfo {author} {\bibfnamefont {T.}~\bibnamefont
  {Taniguchi}},  \emph {et~al.},\ }\href@noop {} {\bibfield  {journal}
  {\bibinfo  {journal} {Nature communications}\ }\textbf {\bibinfo {volume}
  {9}},\ \bibinfo {pages} {1} (\bibinfo {year} {2018})}\BibitemShut {NoStop}%
\bibitem [{\citenamefont {Zhang}\ \emph {et~al.}(2015)\citenamefont {Zhang},
  \citenamefont {Kidd},\ and\ \citenamefont {Varga}}]{zhang2015excited}%
  \BibitemOpen
  \bibfield  {author} {\bibinfo {author} {\bibfnamefont {D.~K.}\ \bibnamefont
  {Zhang}}, \bibinfo {author} {\bibfnamefont {D.~W.}\ \bibnamefont {Kidd}}, \
  and\ \bibinfo {author} {\bibfnamefont {K.}~\bibnamefont {Varga}},\
  }\href@noop {} {\bibfield  {journal} {\bibinfo  {journal} {Nano letters}\
  }\textbf {\bibinfo {volume} {15}},\ \bibinfo {pages} {7002} (\bibinfo {year}
  {2015})}\BibitemShut {NoStop}%
\bibitem [{\citenamefont {Chen}\ \emph {et~al.}(2018)\citenamefont {Chen},
  \citenamefont {Goldstein}, \citenamefont {Taniguchi}, \citenamefont
  {Watanabe},\ and\ \citenamefont {Yan}}]{chen2018coulomb}%
  \BibitemOpen
  \bibfield  {author} {\bibinfo {author} {\bibfnamefont {S.-Y.}\ \bibnamefont
  {Chen}}, \bibinfo {author} {\bibfnamefont {T.}~\bibnamefont {Goldstein}},
  \bibinfo {author} {\bibfnamefont {T.}~\bibnamefont {Taniguchi}}, \bibinfo
  {author} {\bibfnamefont {K.}~\bibnamefont {Watanabe}}, \ and\ \bibinfo
  {author} {\bibfnamefont {J.}~\bibnamefont {Yan}},\ }\href@noop {} {\bibfield
  {journal} {\bibinfo  {journal} {Nature communications}\ }\textbf {\bibinfo
  {volume} {9}},\ \bibinfo {pages} {1} (\bibinfo {year} {2018})}\BibitemShut
  {NoStop}%
\bibitem [{\citenamefont {Liu}\ \emph {et~al.}(2019)\citenamefont {Liu},
  \citenamefont {van Baren}, \citenamefont {Taniguchi}, \citenamefont
  {Watanabe}, \citenamefont {Chang},\ and\ \citenamefont
  {Lui}}]{liu2019valley}%
  \BibitemOpen
  \bibfield  {author} {\bibinfo {author} {\bibfnamefont {E.}~\bibnamefont
  {Liu}}, \bibinfo {author} {\bibfnamefont {J.}~\bibnamefont {van Baren}},
  \bibinfo {author} {\bibfnamefont {T.}~\bibnamefont {Taniguchi}}, \bibinfo
  {author} {\bibfnamefont {K.}~\bibnamefont {Watanabe}}, \bibinfo {author}
  {\bibfnamefont {Y.-C.}\ \bibnamefont {Chang}}, \ and\ \bibinfo {author}
  {\bibfnamefont {C.~H.}\ \bibnamefont {Lui}},\ }\href@noop {} {\bibfield
  {journal} {\bibinfo  {journal} {Physical Review Research}\ }\textbf {\bibinfo
  {volume} {1}},\ \bibinfo {pages} {032007} (\bibinfo {year}
  {2019})}\BibitemShut {NoStop}%
\bibitem [{\citenamefont {Li}\ \emph {et~al.}(2019{\natexlab{a}})\citenamefont
  {Li}, \citenamefont {Wang}, \citenamefont {Jin}, \citenamefont {Lu},
  \citenamefont {Lian}, \citenamefont {Meng}, \citenamefont {Blei},
  \citenamefont {Gao}, \citenamefont {Taniguchi}, \citenamefont {Watanabe}
  \emph {et~al.}}]{li2019emerging}%
  \BibitemOpen
  \bibfield  {author} {\bibinfo {author} {\bibfnamefont {Z.}~\bibnamefont
  {Li}}, \bibinfo {author} {\bibfnamefont {T.}~\bibnamefont {Wang}}, \bibinfo
  {author} {\bibfnamefont {C.}~\bibnamefont {Jin}}, \bibinfo {author}
  {\bibfnamefont {Z.}~\bibnamefont {Lu}}, \bibinfo {author} {\bibfnamefont
  {Z.}~\bibnamefont {Lian}}, \bibinfo {author} {\bibfnamefont {Y.}~\bibnamefont
  {Meng}}, \bibinfo {author} {\bibfnamefont {M.}~\bibnamefont {Blei}}, \bibinfo
  {author} {\bibfnamefont {S.}~\bibnamefont {Gao}}, \bibinfo {author}
  {\bibfnamefont {T.}~\bibnamefont {Taniguchi}}, \bibinfo {author}
  {\bibfnamefont {K.}~\bibnamefont {Watanabe}},  \emph {et~al.},\ }\href@noop
  {} {\bibfield  {journal} {\bibinfo  {journal} {Nature communications}\
  }\textbf {\bibinfo {volume} {10}},\ \bibinfo {pages} {1} (\bibinfo {year}
  {2019}{\natexlab{a}})}\BibitemShut {NoStop}%
\bibitem [{\citenamefont {F{\"o}rste}\ \emph {et~al.}(2020)\citenamefont
  {F{\"o}rste}, \citenamefont {Tepliakov}, \citenamefont {Kruchinin},
  \citenamefont {Lindlau}, \citenamefont {Funk}, \citenamefont {F{\"o}rg},
  \citenamefont {Watanabe}, \citenamefont {Taniguchi}, \citenamefont
  {Baimuratov},\ and\ \citenamefont {H{\"o}gele}}]{forste2020exciton}%
  \BibitemOpen
  \bibfield  {author} {\bibinfo {author} {\bibfnamefont {J.}~\bibnamefont
  {F{\"o}rste}}, \bibinfo {author} {\bibfnamefont {N.~V.}\ \bibnamefont
  {Tepliakov}}, \bibinfo {author} {\bibfnamefont {S.~Y.}\ \bibnamefont
  {Kruchinin}}, \bibinfo {author} {\bibfnamefont {J.}~\bibnamefont {Lindlau}},
  \bibinfo {author} {\bibfnamefont {V.}~\bibnamefont {Funk}}, \bibinfo {author}
  {\bibfnamefont {M.}~\bibnamefont {F{\"o}rg}}, \bibinfo {author}
  {\bibfnamefont {K.}~\bibnamefont {Watanabe}}, \bibinfo {author}
  {\bibfnamefont {T.}~\bibnamefont {Taniguchi}}, \bibinfo {author}
  {\bibfnamefont {A.~S.}\ \bibnamefont {Baimuratov}}, \ and\ \bibinfo {author}
  {\bibfnamefont {A.}~\bibnamefont {H{\"o}gele}},\ }\href@noop {} {\bibfield
  {journal} {\bibinfo  {journal} {Nature Communications}\ }\textbf {\bibinfo
  {volume} {11}},\ \bibinfo {pages} {1} (\bibinfo {year} {2020})}\BibitemShut
  {NoStop}%
\bibitem [{\citenamefont {Steinhoff}\ \emph {et~al.}(2018)\citenamefont
  {Steinhoff}, \citenamefont {Florian}, \citenamefont {Singh}, \citenamefont
  {Tran}, \citenamefont {Kolarczik}, \citenamefont {Helmrich}, \citenamefont
  {Achtstein}, \citenamefont {Woggon}, \citenamefont {Owschimikow},
  \citenamefont {Jahnke} \emph {et~al.}}]{steinhoff2018biexciton}%
  \BibitemOpen
  \bibfield  {author} {\bibinfo {author} {\bibfnamefont {A.}~\bibnamefont
  {Steinhoff}}, \bibinfo {author} {\bibfnamefont {M.}~\bibnamefont {Florian}},
  \bibinfo {author} {\bibfnamefont {A.}~\bibnamefont {Singh}}, \bibinfo
  {author} {\bibfnamefont {K.}~\bibnamefont {Tran}}, \bibinfo {author}
  {\bibfnamefont {M.}~\bibnamefont {Kolarczik}}, \bibinfo {author}
  {\bibfnamefont {S.}~\bibnamefont {Helmrich}}, \bibinfo {author}
  {\bibfnamefont {A.~W.}\ \bibnamefont {Achtstein}}, \bibinfo {author}
  {\bibfnamefont {U.}~\bibnamefont {Woggon}}, \bibinfo {author} {\bibfnamefont
  {N.}~\bibnamefont {Owschimikow}}, \bibinfo {author} {\bibfnamefont
  {F.}~\bibnamefont {Jahnke}},  \emph {et~al.},\ }\href@noop {} {\bibfield
  {journal} {\bibinfo  {journal} {Nature Physics}\ }\textbf {\bibinfo {volume}
  {14}},\ \bibinfo {pages} {1199} (\bibinfo {year} {2018})}\BibitemShut
  {NoStop}%
\bibitem [{\citenamefont {Hao}\ \emph {et~al.}(2017)\citenamefont {Hao},
  \citenamefont {Specht}, \citenamefont {Nagler}, \citenamefont {Xu},
  \citenamefont {Tran}, \citenamefont {Singh}, \citenamefont {Dass},
  \citenamefont {Sch{\"u}ller}, \citenamefont {Korn}, \citenamefont {Richter}
  \emph {et~al.}}]{hao2017neutral}%
  \BibitemOpen
  \bibfield  {author} {\bibinfo {author} {\bibfnamefont {K.}~\bibnamefont
  {Hao}}, \bibinfo {author} {\bibfnamefont {J.~F.}\ \bibnamefont {Specht}},
  \bibinfo {author} {\bibfnamefont {P.}~\bibnamefont {Nagler}}, \bibinfo
  {author} {\bibfnamefont {L.}~\bibnamefont {Xu}}, \bibinfo {author}
  {\bibfnamefont {K.}~\bibnamefont {Tran}}, \bibinfo {author} {\bibfnamefont
  {A.}~\bibnamefont {Singh}}, \bibinfo {author} {\bibfnamefont {C.~K.}\
  \bibnamefont {Dass}}, \bibinfo {author} {\bibfnamefont {C.}~\bibnamefont
  {Sch{\"u}ller}}, \bibinfo {author} {\bibfnamefont {T.}~\bibnamefont {Korn}},
  \bibinfo {author} {\bibfnamefont {M.}~\bibnamefont {Richter}},  \emph
  {et~al.},\ }\href@noop {} {\bibfield  {journal} {\bibinfo  {journal} {Nature
  communications}\ }\textbf {\bibinfo {volume} {8}},\ \bibinfo {pages} {1}
  (\bibinfo {year} {2017})}\BibitemShut {NoStop}%
\bibitem [{\citenamefont {Barbone}\ \emph {et~al.}(2018)\citenamefont
  {Barbone}, \citenamefont {Montblanch}, \citenamefont {Kara}, \citenamefont
  {Palacios-Berraquero}, \citenamefont {Cadore}, \citenamefont {De~Fazio},
  \citenamefont {Pingault}, \citenamefont {Mostaani}, \citenamefont {Li},
  \citenamefont {Chen} \emph {et~al.}}]{barbone2018charge}%
  \BibitemOpen
  \bibfield  {author} {\bibinfo {author} {\bibfnamefont {M.}~\bibnamefont
  {Barbone}}, \bibinfo {author} {\bibfnamefont {A.~R.-P.}\ \bibnamefont
  {Montblanch}}, \bibinfo {author} {\bibfnamefont {D.~M.}\ \bibnamefont
  {Kara}}, \bibinfo {author} {\bibfnamefont {C.}~\bibnamefont
  {Palacios-Berraquero}}, \bibinfo {author} {\bibfnamefont {A.~R.}\
  \bibnamefont {Cadore}}, \bibinfo {author} {\bibfnamefont {D.}~\bibnamefont
  {De~Fazio}}, \bibinfo {author} {\bibfnamefont {B.}~\bibnamefont {Pingault}},
  \bibinfo {author} {\bibfnamefont {E.}~\bibnamefont {Mostaani}}, \bibinfo
  {author} {\bibfnamefont {H.}~\bibnamefont {Li}}, \bibinfo {author}
  {\bibfnamefont {B.}~\bibnamefont {Chen}},  \emph {et~al.},\ }\href@noop {}
  {\bibfield  {journal} {\bibinfo  {journal} {Nature communications}\ }\textbf
  {\bibinfo {volume} {9}},\ \bibinfo {pages} {1} (\bibinfo {year}
  {2018})}\BibitemShut {NoStop}%
\bibitem [{\citenamefont {Debnath}\ \emph {et~al.}(2021)\citenamefont
  {Debnath}, \citenamefont {Sett}, \citenamefont {Biswas}, \citenamefont
  {Raghunathan},\ and\ \citenamefont {Ghosh}}]{debnath2021simple}%
  \BibitemOpen
  \bibfield  {author} {\bibinfo {author} {\bibfnamefont {R.}~\bibnamefont
  {Debnath}}, \bibinfo {author} {\bibfnamefont {S.}~\bibnamefont {Sett}},
  \bibinfo {author} {\bibfnamefont {R.}~\bibnamefont {Biswas}}, \bibinfo
  {author} {\bibfnamefont {V.}~\bibnamefont {Raghunathan}}, \ and\ \bibinfo
  {author} {\bibfnamefont {A.}~\bibnamefont {Ghosh}},\ }\href@noop {}
  {\bibfield  {journal} {\bibinfo  {journal} {Nanotechnology}\ }\textbf
  {\bibinfo {volume} {32}},\ \bibinfo {pages} {455705} (\bibinfo {year}
  {2021})}\BibitemShut {NoStop}%
\bibitem [{\citenamefont {Courtade}\ \emph {et~al.}(2017)\citenamefont
  {Courtade}, \citenamefont {Semina}, \citenamefont {Manca}, \citenamefont
  {Glazov}, \citenamefont {Robert}, \citenamefont {Cadiz}, \citenamefont
  {Wang}, \citenamefont {Taniguchi}, \citenamefont {Watanabe}, \citenamefont
  {Pierre} \emph {et~al.}}]{courtade2017charged}%
  \BibitemOpen
  \bibfield  {author} {\bibinfo {author} {\bibfnamefont {E.}~\bibnamefont
  {Courtade}}, \bibinfo {author} {\bibfnamefont {M.}~\bibnamefont {Semina}},
  \bibinfo {author} {\bibfnamefont {M.}~\bibnamefont {Manca}}, \bibinfo
  {author} {\bibfnamefont {M.}~\bibnamefont {Glazov}}, \bibinfo {author}
  {\bibfnamefont {C.}~\bibnamefont {Robert}}, \bibinfo {author} {\bibfnamefont
  {F.}~\bibnamefont {Cadiz}}, \bibinfo {author} {\bibfnamefont
  {G.}~\bibnamefont {Wang}}, \bibinfo {author} {\bibfnamefont {T.}~\bibnamefont
  {Taniguchi}}, \bibinfo {author} {\bibfnamefont {K.}~\bibnamefont {Watanabe}},
  \bibinfo {author} {\bibfnamefont {M.}~\bibnamefont {Pierre}},  \emph
  {et~al.},\ }\href@noop {} {\bibfield  {journal} {\bibinfo  {journal}
  {Physical Review B}\ }\textbf {\bibinfo {volume} {96}},\ \bibinfo {pages}
  {085302} (\bibinfo {year} {2017})}\BibitemShut {NoStop}%
\bibitem [{\citenamefont {Altaiary}\ \emph {et~al.}(2022)\citenamefont
  {Altaiary}, \citenamefont {Liu}, \citenamefont {Liang}, \citenamefont
  {Hsiao}, \citenamefont {van Baren}, \citenamefont {Taniguchi}, \citenamefont
  {Watanabe}, \citenamefont {Gabor}, \citenamefont {Chang},\ and\ \citenamefont
  {Lui}}]{altaiary2022electrically}%
  \BibitemOpen
  \bibfield  {author} {\bibinfo {author} {\bibfnamefont {M.~M.}\ \bibnamefont
  {Altaiary}}, \bibinfo {author} {\bibfnamefont {E.}~\bibnamefont {Liu}},
  \bibinfo {author} {\bibfnamefont {C.-T.}\ \bibnamefont {Liang}}, \bibinfo
  {author} {\bibfnamefont {F.-C.}\ \bibnamefont {Hsiao}}, \bibinfo {author}
  {\bibfnamefont {J.}~\bibnamefont {van Baren}}, \bibinfo {author}
  {\bibfnamefont {T.}~\bibnamefont {Taniguchi}}, \bibinfo {author}
  {\bibfnamefont {K.}~\bibnamefont {Watanabe}}, \bibinfo {author}
  {\bibfnamefont {N.~M.}\ \bibnamefont {Gabor}}, \bibinfo {author}
  {\bibfnamefont {Y.-C.}\ \bibnamefont {Chang}}, \ and\ \bibinfo {author}
  {\bibfnamefont {C.~H.}\ \bibnamefont {Lui}},\ }\href@noop {} {\bibfield
  {journal} {\bibinfo  {journal} {Nano Letters}\ } (\bibinfo {year}
  {2022})}\BibitemShut {NoStop}%
\bibitem [{\citenamefont {Scuri}\ \emph {et~al.}(2020)\citenamefont {Scuri},
  \citenamefont {Andersen}, \citenamefont {Zhou}, \citenamefont {Wild},
  \citenamefont {Sung}, \citenamefont {Gelly}, \citenamefont {B{\'e}rub{\'e}},
  \citenamefont {Heo}, \citenamefont {Shao}, \citenamefont {Joe} \emph
  {et~al.}}]{scuri2020electrically}%
  \BibitemOpen
  \bibfield  {author} {\bibinfo {author} {\bibfnamefont {G.}~\bibnamefont
  {Scuri}}, \bibinfo {author} {\bibfnamefont {T.~I.}\ \bibnamefont {Andersen}},
  \bibinfo {author} {\bibfnamefont {Y.}~\bibnamefont {Zhou}}, \bibinfo {author}
  {\bibfnamefont {D.~S.}\ \bibnamefont {Wild}}, \bibinfo {author}
  {\bibfnamefont {J.}~\bibnamefont {Sung}}, \bibinfo {author} {\bibfnamefont
  {R.~J.}\ \bibnamefont {Gelly}}, \bibinfo {author} {\bibfnamefont
  {D.}~\bibnamefont {B{\'e}rub{\'e}}}, \bibinfo {author} {\bibfnamefont
  {H.}~\bibnamefont {Heo}}, \bibinfo {author} {\bibfnamefont {L.}~\bibnamefont
  {Shao}}, \bibinfo {author} {\bibfnamefont {A.~Y.}\ \bibnamefont {Joe}},
  \emph {et~al.},\ }\href@noop {} {\bibfield  {journal} {\bibinfo  {journal}
  {Physical Review Letters}\ }\textbf {\bibinfo {volume} {124}},\ \bibinfo
  {pages} {217403} (\bibinfo {year} {2020})}\BibitemShut {NoStop}%
\bibitem [{\citenamefont {Li}\ \emph {et~al.}(2019{\natexlab{b}})\citenamefont
  {Li}, \citenamefont {Wang}, \citenamefont {Jin}, \citenamefont {Lu},
  \citenamefont {Lian}, \citenamefont {Meng}, \citenamefont {Blei},
  \citenamefont {Gao}, \citenamefont {Taniguchi}, \citenamefont {Watanabe}
  \emph {et~al.}}]{li2019momentum}%
  \BibitemOpen
  \bibfield  {author} {\bibinfo {author} {\bibfnamefont {Z.}~\bibnamefont
  {Li}}, \bibinfo {author} {\bibfnamefont {T.}~\bibnamefont {Wang}}, \bibinfo
  {author} {\bibfnamefont {C.}~\bibnamefont {Jin}}, \bibinfo {author}
  {\bibfnamefont {Z.}~\bibnamefont {Lu}}, \bibinfo {author} {\bibfnamefont
  {Z.}~\bibnamefont {Lian}}, \bibinfo {author} {\bibfnamefont {Y.}~\bibnamefont
  {Meng}}, \bibinfo {author} {\bibfnamefont {M.}~\bibnamefont {Blei}}, \bibinfo
  {author} {\bibfnamefont {M.}~\bibnamefont {Gao}}, \bibinfo {author}
  {\bibfnamefont {T.}~\bibnamefont {Taniguchi}}, \bibinfo {author}
  {\bibfnamefont {K.}~\bibnamefont {Watanabe}},  \emph {et~al.},\ }\href@noop
  {} {\bibfield  {journal} {\bibinfo  {journal} {ACS nano}\ }\textbf {\bibinfo
  {volume} {13}},\ \bibinfo {pages} {14107} (\bibinfo {year}
  {2019}{\natexlab{b}})}\BibitemShut {NoStop}%
\bibitem [{\citenamefont {Merkl}\ \emph {et~al.}(2020)\citenamefont {Merkl},
  \citenamefont {Mooshammer}, \citenamefont {Brem}, \citenamefont {Girnghuber},
  \citenamefont {Lin}, \citenamefont {Weigl}, \citenamefont {Liebich},
  \citenamefont {Yong}, \citenamefont {Gillen}, \citenamefont {Maultzsch} \emph
  {et~al.}}]{merkl2020twist}%
  \BibitemOpen
  \bibfield  {author} {\bibinfo {author} {\bibfnamefont {P.}~\bibnamefont
  {Merkl}}, \bibinfo {author} {\bibfnamefont {F.}~\bibnamefont {Mooshammer}},
  \bibinfo {author} {\bibfnamefont {S.}~\bibnamefont {Brem}}, \bibinfo {author}
  {\bibfnamefont {A.}~\bibnamefont {Girnghuber}}, \bibinfo {author}
  {\bibfnamefont {K.-Q.}\ \bibnamefont {Lin}}, \bibinfo {author} {\bibfnamefont
  {L.}~\bibnamefont {Weigl}}, \bibinfo {author} {\bibfnamefont
  {M.}~\bibnamefont {Liebich}}, \bibinfo {author} {\bibfnamefont {C.-K.}\
  \bibnamefont {Yong}}, \bibinfo {author} {\bibfnamefont {R.}~\bibnamefont
  {Gillen}}, \bibinfo {author} {\bibfnamefont {J.}~\bibnamefont {Maultzsch}},
  \emph {et~al.},\ }\href@noop {} {\bibfield  {journal} {\bibinfo  {journal}
  {Nature communications}\ }\textbf {\bibinfo {volume} {11}},\ \bibinfo {pages}
  {1} (\bibinfo {year} {2020})}\BibitemShut {NoStop}%
\bibitem [{\citenamefont {Wang}\ \emph {et~al.}(2018)\citenamefont {Wang},
  \citenamefont {Chiu}, \citenamefont {Honz}, \citenamefont {Mak},\ and\
  \citenamefont {Shan}}]{wang2018electrical}%
  \BibitemOpen
  \bibfield  {author} {\bibinfo {author} {\bibfnamefont {Z.}~\bibnamefont
  {Wang}}, \bibinfo {author} {\bibfnamefont {Y.-H.}\ \bibnamefont {Chiu}},
  \bibinfo {author} {\bibfnamefont {K.}~\bibnamefont {Honz}}, \bibinfo {author}
  {\bibfnamefont {K.~F.}\ \bibnamefont {Mak}}, \ and\ \bibinfo {author}
  {\bibfnamefont {J.}~\bibnamefont {Shan}},\ }\href@noop {} {\bibfield
  {journal} {\bibinfo  {journal} {Nano letters}\ }\textbf {\bibinfo {volume}
  {18}},\ \bibinfo {pages} {137} (\bibinfo {year} {2018})}\BibitemShut
  {NoStop}%
\bibitem [{\citenamefont {Altaiary}\ \emph {et~al.}(2021)\citenamefont
  {Altaiary}, \citenamefont {Liu}, \citenamefont {Liang}, \citenamefont
  {Hsiao}, \citenamefont {van Baren}, \citenamefont {Taniguchi}, \citenamefont
  {Watanabe}, \citenamefont {Gabor}, \citenamefont {Chang},\ and\ \citenamefont
  {Lui}}]{altaiary2021electric}%
  \BibitemOpen
  \bibfield  {author} {\bibinfo {author} {\bibfnamefont {M.~M.}\ \bibnamefont
  {Altaiary}}, \bibinfo {author} {\bibfnamefont {E.}~\bibnamefont {Liu}},
  \bibinfo {author} {\bibfnamefont {C.-T.}\ \bibnamefont {Liang}}, \bibinfo
  {author} {\bibfnamefont {F.-C.}\ \bibnamefont {Hsiao}}, \bibinfo {author}
  {\bibfnamefont {J.}~\bibnamefont {van Baren}}, \bibinfo {author}
  {\bibfnamefont {T.}~\bibnamefont {Taniguchi}}, \bibinfo {author}
  {\bibfnamefont {K.}~\bibnamefont {Watanabe}}, \bibinfo {author}
  {\bibfnamefont {N.~M.}\ \bibnamefont {Gabor}}, \bibinfo {author}
  {\bibfnamefont {Y.-C.}\ \bibnamefont {Chang}}, \ and\ \bibinfo {author}
  {\bibfnamefont {C.~H.}\ \bibnamefont {Lui}},\ }\href@noop {} {\bibfield
  {journal} {\bibinfo  {journal} {arXiv preprint arXiv:2101.11161}\ } (\bibinfo
  {year} {2021})}\BibitemShut {NoStop}%
\bibitem [{\citenamefont {Kyl{\"a}np{\"a}{\"a}}\ and\ \citenamefont
  {Komsa}(2015)}]{kylanpaa2015binding}%
  \BibitemOpen
  \bibfield  {author} {\bibinfo {author} {\bibfnamefont {I.}~\bibnamefont
  {Kyl{\"a}np{\"a}{\"a}}}\ and\ \bibinfo {author} {\bibfnamefont {H.-P.}\
  \bibnamefont {Komsa}},\ }\href@noop {} {\bibfield  {journal} {\bibinfo
  {journal} {Physical Review B}\ }\textbf {\bibinfo {volume} {92}},\ \bibinfo
  {pages} {205418} (\bibinfo {year} {2015})}\BibitemShut {NoStop}%
\bibitem [{\citenamefont {Sung}\ \emph {et~al.}(2020)\citenamefont {Sung},
  \citenamefont {Zhou}, \citenamefont {Scuri}, \citenamefont {Zlyomi},
  \citenamefont {Andersen}, \citenamefont {Yoo}, \citenamefont {Wild},
  \citenamefont {Joe}, \citenamefont {Gelly}, \citenamefont {Heo} \emph
  {et~al.}}]{sung2020broken}%
  \BibitemOpen
  \bibfield  {author} {\bibinfo {author} {\bibfnamefont {J.}~\bibnamefont
  {Sung}}, \bibinfo {author} {\bibfnamefont {Y.}~\bibnamefont {Zhou}}, \bibinfo
  {author} {\bibfnamefont {G.}~\bibnamefont {Scuri}}, \bibinfo {author}
  {\bibfnamefont {V.}~\bibnamefont {Zlyomi}}, \bibinfo {author} {\bibfnamefont
  {T.~I.}\ \bibnamefont {Andersen}}, \bibinfo {author} {\bibfnamefont
  {H.}~\bibnamefont {Yoo}}, \bibinfo {author} {\bibfnamefont {D.~S.}\
  \bibnamefont {Wild}}, \bibinfo {author} {\bibfnamefont {A.~Y.}\ \bibnamefont
  {Joe}}, \bibinfo {author} {\bibfnamefont {R.~J.}\ \bibnamefont {Gelly}},
  \bibinfo {author} {\bibfnamefont {H.}~\bibnamefont {Heo}},  \emph {et~al.},\
  }\href@noop {} {\bibfield  {journal} {\bibinfo  {journal} {Nature
  Nanotechnology}\ }\textbf {\bibinfo {volume} {15}},\ \bibinfo {pages} {750}
  (\bibinfo {year} {2020})}\BibitemShut {NoStop}%
\bibitem [{\citenamefont {Weston}\ \emph {et~al.}(2020)\citenamefont {Weston},
  \citenamefont {Zou}, \citenamefont {Enaldiev}, \citenamefont {Summerfield},
  \citenamefont {Clark}, \citenamefont {Z{\'o}lyomi}, \citenamefont {Graham},
  \citenamefont {Yelgel}, \citenamefont {Magorrian}, \citenamefont {Zhou} \emph
  {et~al.}}]{weston2020atomic}%
  \BibitemOpen
  \bibfield  {author} {\bibinfo {author} {\bibfnamefont {A.}~\bibnamefont
  {Weston}}, \bibinfo {author} {\bibfnamefont {Y.}~\bibnamefont {Zou}},
  \bibinfo {author} {\bibfnamefont {V.}~\bibnamefont {Enaldiev}}, \bibinfo
  {author} {\bibfnamefont {A.}~\bibnamefont {Summerfield}}, \bibinfo {author}
  {\bibfnamefont {N.}~\bibnamefont {Clark}}, \bibinfo {author} {\bibfnamefont
  {V.}~\bibnamefont {Z{\'o}lyomi}}, \bibinfo {author} {\bibfnamefont
  {A.}~\bibnamefont {Graham}}, \bibinfo {author} {\bibfnamefont
  {C.}~\bibnamefont {Yelgel}}, \bibinfo {author} {\bibfnamefont
  {S.}~\bibnamefont {Magorrian}}, \bibinfo {author} {\bibfnamefont
  {M.}~\bibnamefont {Zhou}},  \emph {et~al.},\ }\href@noop {} {\bibfield
  {journal} {\bibinfo  {journal} {Nature Nanotechnology}\ }\textbf {\bibinfo
  {volume} {15}},\ \bibinfo {pages} {592} (\bibinfo {year} {2020})}\BibitemShut
  {NoStop}%
\bibitem [{\citenamefont {Pei}\ \emph {et~al.}(2015)\citenamefont {Pei},
  \citenamefont {Yang}, \citenamefont {Xu}, \citenamefont {Zeng}, \citenamefont
  {Myint}, \citenamefont {Zhang}, \citenamefont {Zheng}, \citenamefont {Qin},
  \citenamefont {Wang}, \citenamefont {Jiang} \emph {et~al.}}]{pei2015exciton}%
  \BibitemOpen
  \bibfield  {author} {\bibinfo {author} {\bibfnamefont {J.}~\bibnamefont
  {Pei}}, \bibinfo {author} {\bibfnamefont {J.}~\bibnamefont {Yang}}, \bibinfo
  {author} {\bibfnamefont {R.}~\bibnamefont {Xu}}, \bibinfo {author}
  {\bibfnamefont {Y.-H.}\ \bibnamefont {Zeng}}, \bibinfo {author}
  {\bibfnamefont {Y.~W.}\ \bibnamefont {Myint}}, \bibinfo {author}
  {\bibfnamefont {S.}~\bibnamefont {Zhang}}, \bibinfo {author} {\bibfnamefont
  {J.-C.}\ \bibnamefont {Zheng}}, \bibinfo {author} {\bibfnamefont
  {Q.}~\bibnamefont {Qin}}, \bibinfo {author} {\bibfnamefont {X.}~\bibnamefont
  {Wang}}, \bibinfo {author} {\bibfnamefont {W.}~\bibnamefont {Jiang}},  \emph
  {et~al.},\ }\href@noop {} {\bibfield  {journal} {\bibinfo  {journal} {Small}\
  }\textbf {\bibinfo {volume} {11}},\ \bibinfo {pages} {6384} (\bibinfo {year}
  {2015})}\BibitemShut {NoStop}%
\bibitem [{\citenamefont {Shibata}(1998)}]{shibata1998negative}%
  \BibitemOpen
  \bibfield  {author} {\bibinfo {author} {\bibfnamefont {H.}~\bibnamefont
  {Shibata}},\ }\href@noop {} {\bibfield  {journal} {\bibinfo  {journal}
  {Japanese journal of applied physics}\ }\textbf {\bibinfo {volume} {37}},\
  \bibinfo {pages} {550} (\bibinfo {year} {1998})}\BibitemShut {NoStop}%
\bibitem [{\citenamefont {Kaindl}\ \emph {et~al.}(2009)\citenamefont {Kaindl},
  \citenamefont {H{\"a}gele}, \citenamefont {Carnahan},\ and\ \citenamefont
  {Chemla}}]{kaindl2009transient}%
  \BibitemOpen
  \bibfield  {author} {\bibinfo {author} {\bibfnamefont {R.~A.}\ \bibnamefont
  {Kaindl}}, \bibinfo {author} {\bibfnamefont {D.}~\bibnamefont {H{\"a}gele}},
  \bibinfo {author} {\bibfnamefont {M.}~\bibnamefont {Carnahan}}, \ and\
  \bibinfo {author} {\bibfnamefont {D.}~\bibnamefont {Chemla}},\ }\href@noop {}
  {\bibfield  {journal} {\bibinfo  {journal} {Physical Review B}\ }\textbf
  {\bibinfo {volume} {79}},\ \bibinfo {pages} {045320} (\bibinfo {year}
  {2009})}\BibitemShut {NoStop}%
\bibitem [{\citenamefont {Mock}\ \emph {et~al.}(1978)\citenamefont {Mock},
  \citenamefont {Thomas},\ and\ \citenamefont {Combescot}}]{mock1978entropy}%
  \BibitemOpen
  \bibfield  {author} {\bibinfo {author} {\bibfnamefont {J.}~\bibnamefont
  {Mock}}, \bibinfo {author} {\bibfnamefont {G.}~\bibnamefont {Thomas}}, \ and\
  \bibinfo {author} {\bibfnamefont {M.}~\bibnamefont {Combescot}},\ }\href@noop
  {} {\bibfield  {journal} {\bibinfo  {journal} {Solid State Communications}\
  }\textbf {\bibinfo {volume} {25}},\ \bibinfo {pages} {279} (\bibinfo {year}
  {1978})}\BibitemShut {NoStop}%
\bibitem [{\citenamefont {Huang}\ \emph {et~al.}(2016)\citenamefont {Huang},
  \citenamefont {Hoang},\ and\ \citenamefont {Mikkelsen}}]{huang2016probing}%
  \BibitemOpen
  \bibfield  {author} {\bibinfo {author} {\bibfnamefont {J.}~\bibnamefont
  {Huang}}, \bibinfo {author} {\bibfnamefont {T.~B.}\ \bibnamefont {Hoang}}, \
  and\ \bibinfo {author} {\bibfnamefont {M.~H.}\ \bibnamefont {Mikkelsen}},\
  }\href@noop {} {\bibfield  {journal} {\bibinfo  {journal} {Scientific
  reports}\ }\textbf {\bibinfo {volume} {6}},\ \bibinfo {pages} {1} (\bibinfo
  {year} {2016})}\BibitemShut {NoStop}%
\bibitem [{\citenamefont {Giannozzi}\ \emph {et~al.}(2009)\citenamefont
  {Giannozzi}, \citenamefont {Baroni}, \citenamefont {Bonini}, \citenamefont
  {Calandra}, \citenamefont {Car}, \citenamefont {Cavazzoni}, \citenamefont
  {Ceresoli}, \citenamefont {Chiarotti}, \citenamefont {Cococcioni},
  \citenamefont {Dabo} \emph {et~al.}}]{giannozzi2009quantum}%
  \BibitemOpen
  \bibfield  {author} {\bibinfo {author} {\bibfnamefont {P.}~\bibnamefont
  {Giannozzi}}, \bibinfo {author} {\bibfnamefont {S.}~\bibnamefont {Baroni}},
  \bibinfo {author} {\bibfnamefont {N.}~\bibnamefont {Bonini}}, \bibinfo
  {author} {\bibfnamefont {M.}~\bibnamefont {Calandra}}, \bibinfo {author}
  {\bibfnamefont {R.}~\bibnamefont {Car}}, \bibinfo {author} {\bibfnamefont
  {C.}~\bibnamefont {Cavazzoni}}, \bibinfo {author} {\bibfnamefont
  {D.}~\bibnamefont {Ceresoli}}, \bibinfo {author} {\bibfnamefont {G.~L.}\
  \bibnamefont {Chiarotti}}, \bibinfo {author} {\bibfnamefont {M.}~\bibnamefont
  {Cococcioni}}, \bibinfo {author} {\bibfnamefont {I.}~\bibnamefont {Dabo}},
  \emph {et~al.},\ }\href@noop {} {\bibfield  {journal} {\bibinfo  {journal}
  {Journal of physics: Condensed matter}\ }\textbf {\bibinfo {volume} {21}},\
  \bibinfo {pages} {395502} (\bibinfo {year} {2009})}\BibitemShut {NoStop}%
\bibitem [{\citenamefont {Perdew}\ and\ \citenamefont
  {Zunger}(1981)}]{perdew1981self}%
  \BibitemOpen
  \bibfield  {author} {\bibinfo {author} {\bibfnamefont {J.~P.}\ \bibnamefont
  {Perdew}}\ and\ \bibinfo {author} {\bibfnamefont {A.}~\bibnamefont
  {Zunger}},\ }\href@noop {} {\bibfield  {journal} {\bibinfo  {journal}
  {Physical Review B}\ }\textbf {\bibinfo {volume} {23}},\ \bibinfo {pages}
  {5048} (\bibinfo {year} {1981})}\BibitemShut {NoStop}%
\bibitem [{\citenamefont {Monkhorst}\ and\ \citenamefont
  {Pack}(1976)}]{monkhorst1976special}%
  \BibitemOpen
  \bibfield  {author} {\bibinfo {author} {\bibfnamefont {H.~J.}\ \bibnamefont
  {Monkhorst}}\ and\ \bibinfo {author} {\bibfnamefont {J.~D.}\ \bibnamefont
  {Pack}},\ }\href@noop {} {\bibfield  {journal} {\bibinfo  {journal} {Physical
  review B}\ }\textbf {\bibinfo {volume} {13}},\ \bibinfo {pages} {5188}
  (\bibinfo {year} {1976})}\BibitemShut {NoStop}%
\bibitem [{\citenamefont {Popescu}\ and\ \citenamefont
  {Zunger}(2012)}]{popescu2012extracting}%
  \BibitemOpen
  \bibfield  {author} {\bibinfo {author} {\bibfnamefont {V.}~\bibnamefont
  {Popescu}}\ and\ \bibinfo {author} {\bibfnamefont {A.}~\bibnamefont
  {Zunger}},\ }\href@noop {} {\bibfield  {journal} {\bibinfo  {journal}
  {Physical Review B}\ }\textbf {\bibinfo {volume} {85}},\ \bibinfo {pages}
  {085201} (\bibinfo {year} {2012})}\BibitemShut {NoStop}%
\end{thebibliography}%
\bibliographystyle{apsrev4-1}
\newpage
\end{document}